\documentclass{optica-article}

\articletype{Research Article}

\usepackage{float}
\usepackage{graphicx}
\usepackage{amsfonts}
\usepackage{amsmath}
\usepackage{lineno}
\usepackage{caption, subcaption}
\usepackage{bm}
\usepackage{multirow}
\usepackage{soul}

\newcommand{\changes}{\textcolor{black}}

\begin{document}

\title{Learning Robust Features for Scatter Removal and Reconstruction in Dynamic ICF X-Ray Tomography}

\author{Siddhant~Gautam,\authormark{1,*}
Marc~L.~Klasky,\authormark{2}
Balasubramanya T. Nadiga,\authormark{2}
Trevor Wilcox,\authormark{3}
Gary Salazar,\authormark{4}
and Saiprasad~Ravishankar\authormark{1,5}}

\address{ \authormark{1,*} Department of Computational Mathematics, Science and Engineering, Michigan State University, East Lansing, MI, 48824, USA\\
\authormark{2} Theoretical Division, Los Alamos National Laboratory, Los Alamos, NM, 87545, USA\\
\authormark{3,} Theoretical Design Division, Los Alamos National Laboratory, Los Alamos, NM, 87545, USA\\
\authormark{4} J-4 DARHT Experiments \& Radiographic Science, Los Alamos National Laboratory, Los Alamos, NM, 87545, USA\\
\authormark{1,5} Department of Biomedical Engineering, Michigan State University, East Lansing, MI, 48824 USA}

\email{\authormark{*}gautamsi@msu.edu} 
\email{\authormark{\textdagger}ravisha3@msu.edu}

\begin{abstract*} 
Density reconstruction from X-ray projections is an important problem in radiography with key applications in scientific and industrial X-ray computed tomography (CT). Often, such projections are corrupted by unknown sources of noise and scatter, which, when not properly accounted for, can lead to significant errors in density reconstruction. In the setting of this problem, recent deep learning-based methods have shown promise in improving the accuracy of density reconstruction. In this article, we propose a deep learning-based encoder-decoder framework wherein the encoder extracts robust features from noisy/corrupted X-ray projections and the decoder reconstructs the density field from the features extracted by the encoder. We explore three options for the latent-space representation of features: physics-inspired supervision, self-supervision, and no supervision. We find that variants based on self-supervised and physics-inspired supervised features perform better over a range of unknown scatter and noise. In extreme noise settings, the variant with self-supervised features performs best. After investigating further details of the proposed deep-learning methods, we conclude by demonstrating that the newly proposed methods are able to achieve higher accuracy in density reconstruction when compared to a traditional iterative technique.
\end{abstract*}

\section{Introduction}

\begin{figure}[!t]
    \centering
    \includegraphics[width=0.7\linewidth]{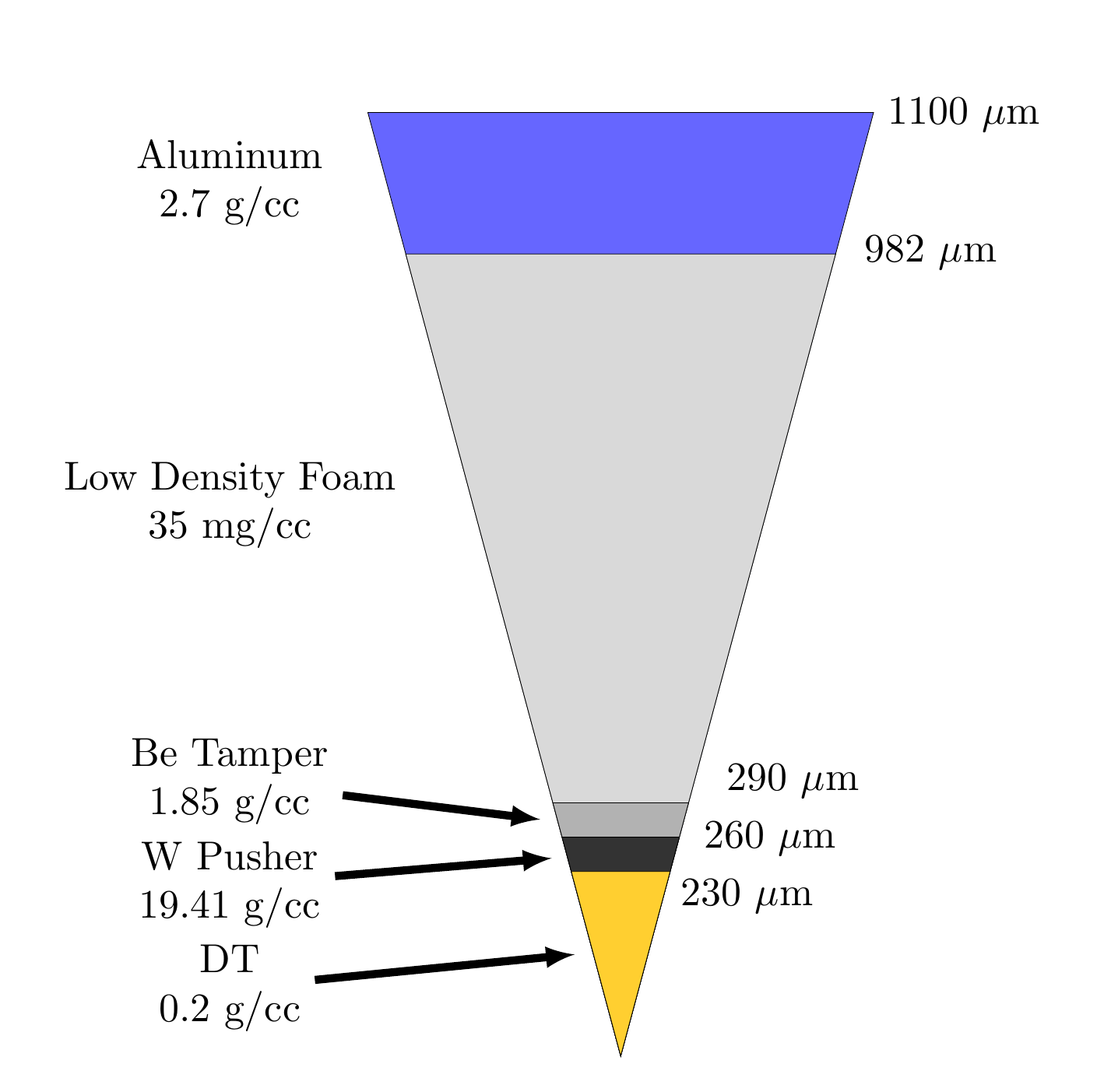}
    \caption{A typical ICF double shell configuration~\cite{merritt2019experimental}.}
    \label{fig:typical_double_shell}
\end{figure}

Radiography plays an important role in various scientific domains such as materials science, shock physics, and inertial confinement fusion (ICF) applications. In these different domains, radiography is used to probe evolving density fields and to understand fundamental physics underlying various associated phenomena. In this work, we choose a setup {comprising} a double-shell ICF capsule (Figure \ref{fig:typical_double_shell}) that employs a high Z metallic shell that implodes onto a gas-filled cavity.
In such a setup, the growth of hydrodynamic instabilities due to geometric and drive asymmetries adversely impacts the level of compression achieved by the implosion, necessitating further investigation of the impact of such asymmetries. The problem setup illustrates the role radiography plays in capturing the complex behavior of various types of material interfaces that are rapidly evolving. 

Material density reconstruction from such X-ray projections has been a challenging task, and some of the analytical image reconstruction methods in computed tomography (CT) range from filtered-back projection (FBP) methods, to the FDK (Feldkamp, Davis and Kress) method and the Inverse Abel Transform~\cite{cormack1963representation, feldkamp1984practical, bracewell1986fourier}. The intricate and noisy nature of the multi-scale, multi-physics environments in many applications still presents a formidable challenge to achieving accurate density field reconstructions. 

 There has been significant progress in the development of iterative reconstruction techniques that incorporate the physics of the problem as well as measurement and noise statistics.
 These methods are often referred to as model-based iterative reconstruction (MBIR) methods~\cite{elbakri2002statistical, ravishankar2019image}. The image reconstruction problem for these MBIR methods takes the form of the following optimization problem:
\begin{align}
   \underset{\mathbf{x} \in \mathcal{D} }{\min}\, F(\mathbf{x},\mathbf{y}) + \alpha  \mathcal{R}(\mathbf{x})
   \label{eq:forward_model}
\end{align}
where $F$ is the data fidelity term capturing the forward operator of the imaging process, $\mathbf{x}$ is the reconstructed density or attenuation map, $\mathbf{y}$ denotes the noisy measurements, and $\mathcal{D}$ captures constraints such as non-negativity. $\mathcal{R}(\cdot)$ is a prior that enforces assumed properties on the reconstructed object and $\alpha$ is a parameter controlling the relative weighting of the two terms. 

 While techniques for tomographic reconstruction, like filtered back projection or model-based iterative reconstruction (MBIR), are accessible, significant challenges persist in accurately determining the material composition of the object. Similar difficulties are also encountered in contexts like baggage screening and nondestructive testing. As a result, these methods often yield distorted images in the presence of X-ray scatter, limited views, etc. Furthermore, these effects may result in a reduction in image contrast and the emergence of image anomalies such as cupping, shading, streaks, and more. The scattering may be attributable to various interactions between photons and matter~\cite{cohen1998atom}, which encompass Compton scatter~\cite{compton1923quantum}, Rayleigh scatter~\cite{young1981rayleigh}, pair production, as well as scatter events involving the scene or background. However, great difficulty is typically encountered in developing accurate models of scatter and noise in practice.

Numerous methodologies have been put forth to address scatter correction~\cite{stonestrom1976scatter, sun2010improved, bhatia2017convolution, tisseur2018evaluation} in domains like medical imaging, nondestructive testing, and related contexts~\cite{ruhrnschopf2011general, ruhrnschopf2011generalpart2}. Contemporary approaches~\cite{maier2018deep,mccann2021local} involve scatter correction through learning or fitting by utilizing training datasets that incorporate examples generated via Monte Carlo N-particle transport code (MCNP) simulations~\cite{werner2018mcnp}. Other recent works attempt to remove scatter and noise by denoising corrupted densities~\cite{huang2022physics} using generative adversarial networks (GANs) or by exploring physical image features that are unaffected by scatter~\cite{hossain2022high}. 

In this work, we attempt to learn radiographic features that can enable robust density estimates in the presence of scatter and noise in inertial confinement fusion applications. We use a 3D cone beam CT setup and sophisticated perturbation modeling, including blur, scatter, and background fields, to generate synthetic radiographs that better represent realistic setups. The results indicate that our density reconstruction approaches are robust to the presence of large unknown scatter fields and noisy perturbations. Although a feature-based approach has been previously developed~\cite{hossain2022high}, it was demonstrated in a simplified 1D geometry. Furthermore, extensions to this feature-based approach have been proposed, but the features were not directly learned from the noisy synthetic radiograph~\cite{serino2023density}.

The remainder of this paper is organized as follows. In Section~\ref{sec:dynamic_radiograpy}, we discuss the forward model describing the physics of the imaging process and detail the synthetic radiograph generation. We also include the details of noise models used to add noise and other perturbations to our clean synthetic radiographs. Section~\ref{sec:proposed_approach} presents our encoder-decoder-based robust feature extraction and density reconstruction model. In Section~\ref{sec:experiments}, we present numerical results and analysis, and finally, we conclude in Section~\ref{sec:conclusion}. The code for the proposed framework is available on GitHub~\cite{gautam2025dynamic}.

\section{Methodology} \label{sec:dynamic_radiograpy}
\subsection{Dynamic Radiography}
\begin{figure}[H]
    \centering
    \includegraphics[width=0.7\linewidth]{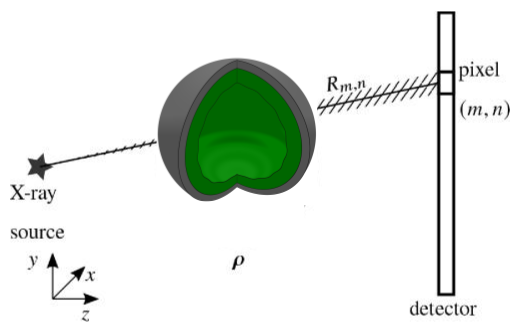}
    \caption{A typical sketch~\cite{mccann2021local} of X-ray tomography where a 3D object $\bm{\rho} \in \mathbb{R}^{N_1\times N_2 \times N_3}$ is being imaged at the detector plane by the set of rays $R_{m,n}$.}
    \label{fig:xray_setup}
\end{figure}

In a typical monoenergetic X-ray tomography setup, {an object is being imaged by an X-ray source at the detector plane as shown in Figure~\ref{fig:xray_setup}}. The X-rays get attenuated by the object, and the amount of attenuation along ray $r$ is proportional to the areal density $\bm{\rho}_A(r)$ {which is given by:}

\begin{equation} 
    \bm{\rho}_A(r) = \int_{-\infty}^\infty \bm{\rho}(r_x(t), r_y(t), r_z(t)) dt,
    \label{eq:areal}
\end{equation}
where $\bm{\rho}$ is the density of the underlying object and $r_x(t), r_y(t), r_z(t)$ denotes the {Cartesian coordinates of the ray $r$ with parameter $t$}.

For a monoenergetic X-ray source and a single material in the object, the number density of photons reaching the detector along ray $r$ is given as:
\begin{equation}
    I(r) = I_0\, exp(-\xi \bm{\rho}_A(r)),
    \label{eq:intensity}
\end{equation}
where $I_0$ is the number density of the incident beam, $\bm{\rho}_A(r)$ is the areal density of the object along ray $r$, and $\xi$ is the mass attenuation coefficient of the material of the object being imaged~\cite{berger2010xcom}.

Sometimes the X-ray beam may pass through more than one material (e.g., gas, metal) before hitting the detector plane. In that case, the radiograph should incorporate the attenuation effects from all materials. In addition, a collimator may also be present in the imaging system to reduce the dynamic range of X-rays {and} reduce the scattered radiation. {The intensity of X-rays reaching the detector}, referred to as the direct radiation or the clean radiograph $\mathbf{D}$, at each pixel, may be written as

\begin{equation}
{ \mathbf{D}_{mn}=\int_{R_{m,n}} I(r) \,dr \approx C \, \exp \left( - \sum_{i=1}^N \xi_i \bm{\rho}_A^i(r_{mn}) \right),}
 \label{eq:direct_radiograph}
 \end{equation}

where $\xi_i$ and $\bm{\rho}_A^i$ are the mass attenuation coefficient and areal mass of the $i$th material, respectively. {$N$ is the number of materials, $r_{mn}$ denotes the ray through the center of pixel $(m, n)$ in the detector, $R_{m,n}$ denotes the set of rays impinging pixel $(m, n)$ and constant $C$ depends on the incident angle and detector pixel area.}
The direct radiation $\mathbf{D}$ may be further corrupted by scatter, noise, and other perturbations, and the measured transmission or the noisy radiograph $\mathbf{T}$ is the sum of the direct radiograph and these perturbations. {The aim of this work} to recover the discretized version of the {underlying} continuous density denoted as $\bm{\rho}\in  \mathbb{R}^{N_1\times N_2 \times N_3}$ from this transmission $\mathbf{T} \in \mathbb{R}^{M_1\times M_2 \times M_v}$. Here, $N_1$, $N_2$, and $N_3$ denote the spatial dimensions of the 3D densities to be reconstructed, and $M_1$ and $M_2$ are the spatial dimensions of the corrupted X-ray projections, and $M_v$ is the number of views. For an axis-symmetric object, one view is sufficient if the variability of stochastic fields, e.g., scatter and noise, for each view is ignored.
{The evolution of the density of the object in a dynamic experiment over time} is governed by physical laws that are described by a system of partial differential equations (PDEs) describing radiation hydrodynamics. To facilitate the analysis of the radiation-hydrodynamic system, we utilize the Euler equations in a manner analogous to Bello-Maldonado~\cite{toro2013riemann, bello2020matrix} as follows:

\begin{equation}
    \partial_t{\rho} + \nabla\cdot(\mathbf{u}\,{\rho}) = 0,\quad {\rho}\,(\partial_t\mathbf{u} + (\mathbf{u}\cdot\nabla)\mathbf{u}) = -\nabla p,\;\; {\rho}\,(\partial_t e  +\mathbf{u}\cdot\nabla e)=-p\,\nabla\cdot \mathbf{u},
    \label{eq:rho_pde}
\end{equation}
where $\rho$ is the density written as a scalar function of spatial coordinates and time, $\mathbf{u}$ is the fluid velocity, $e$ is the specific internal energy, and $p = P(\rho,e)$ is the pressure which is a function of $\rho$ and $e$. The time evolution of the hydrodynamic densities can be found using the Euler equations when the non-dissipative fluid's equation of state (EOS)~\cite{robinson2019mie} and its suitable initial and boundary conditions are known. This simplification to the radiation-hydrodynamic equations has been previously utilized by other researchers~\cite{bello2020matrix}.

\subsection{Scatter, Blur, and Noise Models}
In our studies, we simulate realistic radiographic measurements by introducing source blur, detector blur, correlated scatter, background scatter, and gamma and photon noise to the direct radiographs. Furthermore, we include all these aspects {while testing our trained deep-learning models}, with random variations to mimic model mismatches and stochastic variations. {The details of these perturbations are as follows:}

\begin{enumerate}

\item We model source blur using a 2D Gaussian kernel $\mathbf{G}_{blur}$ with standard deviation $\sigma_{blur}$ in our numerical investigations. The detector blur is modeled using another kernel $\bm{\phi}_{db}$. The result after source and detector blur (denoted as $\mathbf{D}_{blur}$) is obtained by convolving (denoted by `$\ast$') the direct radiation (or the clean radiograph) $\mathbf{D}$ with these kernels as follows:
\begin{equation}
 \mathbf{D}_{blur} = (\mathbf{D} \ast \mathbf{G}_{blur} (\sigma_{blur})) \ast  \bm{\phi}_{db}.
 \label{eq:detector_source_blur}
\end{equation}

\item We model correlated X-ray scatter
via convolution between the direct radiograph and a scatter kernel.
We use a 2D Gaussian filter $\mathbf{G}_{scatter}$ having standard deviation $\sigma_{scatter}$ for the kernel. The correlated scatter denoted as $\mathbf{D}_{s}$ is obtained as follows:
\begin{equation}
 \mathbf{D}_s = \kappa \mathbf{D} \ast \mathbf{G}_{scatter} (\sigma_{scatter}),
 \label{eq:scatter}
\end{equation}
where $\kappa$ is an additional scaling factor.

\item We also added a background scatter field $\mathbf{B}_s$, which is another essential component of scatter affecting radiographic measurements. This field is modeled by a polynomial of order $n$ given as
\begin{equation}
 \mathbf{B}_s = \sum_{i=1}^n a_i x^i + b_iy^i,
 \label{eq:background_scatter}
\end{equation}
where $x$ and $y$ denote spatial coordinates and $a_i$ and $b_i$ are the coefficients of the polynomial. 

\item Gamma and photon noise are modeled as Poisson noise denoted as $\bm{\eta}^{Po}$ {having means $\gamma_g$ and $\gamma_p$} that are proportional to the total signal $\mathbf{D}_{blur}+\mathbf{D}_s+\mathbf{B}_s$ ({with a scaling factor involved for each mean}). The noise components are convolved with respective kernels $\bm{\phi}_g$ and $\bm{\phi}_p$ to give the total (colored) noise $\bm{\eta}$ as follows:
\begin{equation}
{\bm{\eta}=\kappa_g[\bm{\eta}^{Po}(\gamma_g) \ast \bm{\phi}_g] + \kappa_p [\bm{\eta}^{Po}(\gamma_p) \ast \bm{\phi}_p],}
\label{eq:poisson_noise}
\end{equation}

where $\kappa_g$ and $\kappa_p$ are scaling coefficients for the gamma and photon noise components, respectively.
\end{enumerate}

The total transmission (or the noisy radiograph) is the sum of the blurred radiograph, scatter, and noise and is given by:
\begin{equation}
    \mathbf{T} = \mathbf{D}_{blur} + \mathbf{D}_s + \mathbf{B}_s + \bm{\eta}.
    \label{eq:noisy_radiograph}
\end{equation}

\section{Proposed Robust Feature Extraction Based Approach} \label{sec:proposed_approach}
\begin{figure}
    \centering
    \includegraphics[width=0.8\linewidth]{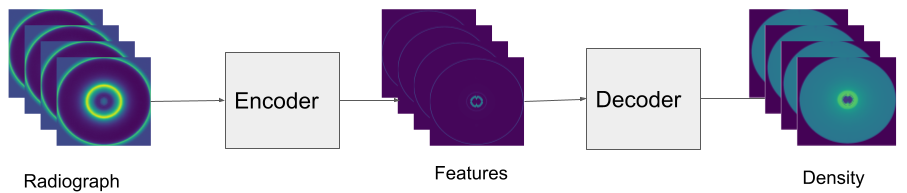}
    \caption{Block diagram of encoder-decoder based architecture. Four different frames of a temporal sequence are used for training.}
    \label{fig:encoder_decoder_stack}
\end{figure}

Our key innovation is to reconstruct density fields using robust features extracted from corrupted radiographs. In this section, we discuss {this approach in which we use} an encoder-decoder-based framework in conjunction with certain constraints. Figure~\ref{fig:encoder_decoder_stack} shows the block diagram of our encoder-decoder framework for extracting features and performing density reconstruction. 
Encoders-decoders belong to a category of neural network architectures that are designed to transform data points from input images to output images by employing a two-stage network~\cite{badrinarayanan2017segnet, goodfellow2016deep}. 

\changes{In this paper, we present three such encoder-decoder-based robust feature extraction and density reconstruction techniques called the physics-inspired supervised latent representation (PISLR), self-supervised latent representation (SSLR), and unsupervised latent representation (ULR) approaches. We note that all three approaches are supervised as far as the final output/target (density) is concerned. When we distinguish between the three variants: PISLR, SSLR, and ULR, we are only referring to how the latent space features are learned and whether there is any supervision involved in that; the final target (i.e., clean density) is always supervised. These are presented in the following subsections.}

\subsection{PISLR Approach}
In earlier work for hydrodynamic density reconstruction, {the shock and edge features of the density profiles} were used in conjunction with the learned dynamics of the evolution of 1D densities~\cite{hossain2022high}. In addition, assumptions regarding the ability to obtain these features from radiographic projections were made for a 1D case. For 2D hydrodynamic simulations (corresponding to 3D axisymmetric objects) that are being examined in this work, it has been found that the recovery of these features in the vicinity of the gas-metal interface is inaccurate. This is due to the inability to locate these features accurately in the presence of noise as well as their unstable nature. Accordingly, to eliminate such sources of errors, we use the edgemaps of the clean density, in which the gas-metal interface is masked out, as the features for training our network and {use these features} to predict the density using the decoder network. We call this approach the physics-inspired supervised latent representation (PISLR) approach, where the loss function for training the encoder-decoder can be represented as

\begin{equation}
    \underset{\bm{\theta}_1, \bm{\theta}_2}{\min} \,\,  \mathbb{E}_{(\bm{\rho},\,\mathbf{T})} \frac{\|D_{\bm{\theta}_2}(E_{\bm{\theta}_1}(\mathbf{T}))- \bm{\rho}\|_2}{\| \bm{\rho}\|_2} + \lambda_{\text{PISLR}} \frac{\|E_{\bm{\theta}_1}(\mathbf{T}) - \mathbf{M} \odot E_f(\bm{\rho})\|_1}{\|\mathbf{M} \odot E_f(\bm{\rho})\|_1}.   
    \label{eq:masked}
\end{equation}

\begin{figure}
    \centering
    \includegraphics[width=0.8\linewidth]{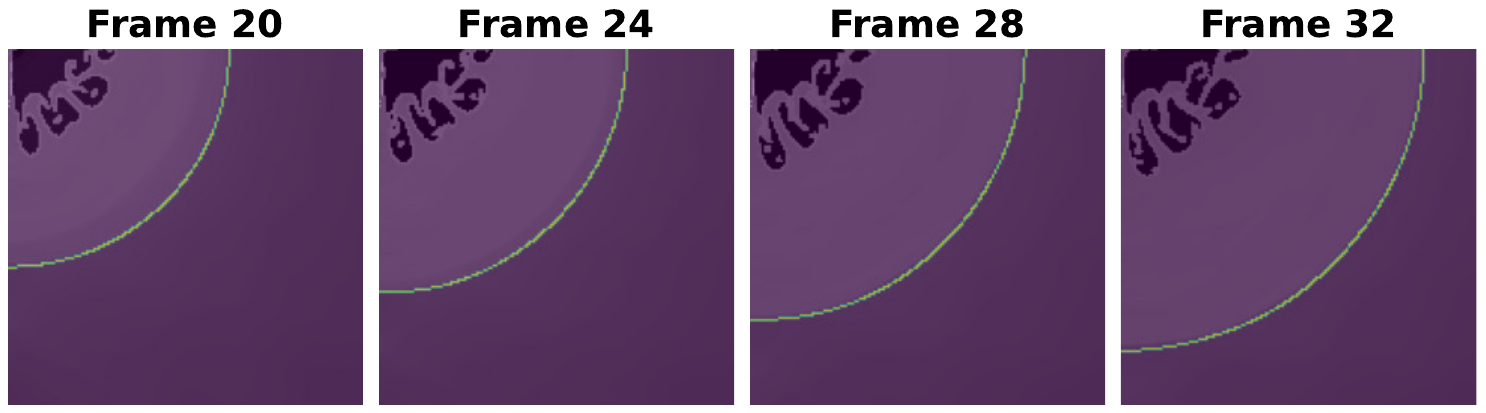}
    \caption{2D profile of the shock edge (indicated by green curve) and the density evolution across four different time instants in the hydrodynamic simulation. Only a quarter of the density profile is shown.}
    \label{fig:shock_2d}
\end{figure}

\begin{figure}
    \centering
    \includegraphics[width=0.8\linewidth]{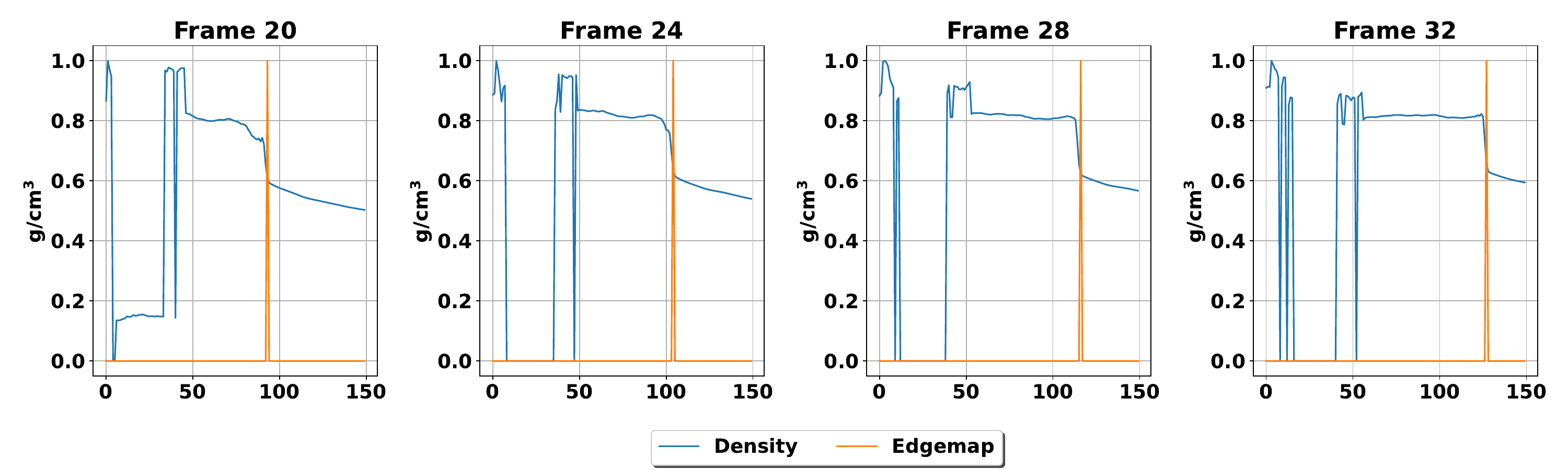}
    \caption{1D line-outs of the shock edge and the density evolution across four different time instants in the simulation.}
    \label{fig:shock_1d}
\end{figure}

Here, $E_{\bm{\theta}_1}$ and $D_{\bm{\theta}_2}$ are the encoder and decoder networks with parameters $\bm{\theta}_1$ and $\bm{\theta}_2$, respectively. $\mathbf{D}$ and $\mathbf{T}$ are the clean and noisy radiographs, respectively, and $\bm{\rho}$ is the underlying clean density that leads to the clean radiograph. $\lambda_{\text{PISLR}}$ is the hyperparameter controlling the weighting of the two terms. The expectation in the loss above is with respect to the distribution of the densities and radiographs.
The first term in the loss function in Eq.~\eqref{eq:masked} captures the error in reconstructing the density fields, and the second term captures the ability of the encoder to capture the edge features. 
\changes{The physical features selected for this PISLR method are the shock features of the density profile, which were found to be robust to scatter and noise in earlier work~\cite{hossain2021high}. In our simulations, these features, denoted by $E_f(\bm{\rho})$, are obtained by first applying a Canny edge filter~\cite{canny1986computational} to the density profiles $\bm{\rho}$, which gives a detailed edgemap containing all edges in the image. A masking operation $\mathbf{M}$ on the edgemap eliminates non-essential edges (e.g., the gas-metal interface) to produce only the outgoing shock features. Figures~\ref{fig:shock_2d} and~\ref{fig:shock_1d} show an example of the 2D profile and 1D line out of the outgoing shock features, evolving over time. These shock features are then used as the labels for training the encoder network (second term in eq.~\eqref{eq:masked}) and hence we name this the physics-inspired supervised latent representation approach. At testing/inference time, the learned encoder $E_{\bm{\theta}_1}$ is applied to the noisy radiographs $\mathbf{T}$ to produce features that closely resemble the original shock features. Thereafter, the learned decoder network is applied to these features to predict the final reconstructed density.}

\subsection{SSLR Approach}
One of the key objectives of our proposed approach (similar to the 1D setting~\cite{hossain2022high}) is to learn and predict features that are robust to unknown noise {and} scatter perturbations. This is because we want to build a robust model that performs equally well even if the noise has characteristics different from those assumed in training the model (unlike earlier works using WGANs~\cite{huang2022physics}). The main idea is to examine learned features that are not limited to shock and edge locations, i.e., to examine the possibility of learning additional or different robust features over the whole image. To examine this, we propose a new approach in which the features are learned {in a self-supervised manner; we call this the self-supervised latent representation (SSLR) approach (as opposed to the PISLR approach)}. The regularization term in this loss function (Eq.~\eqref{eq:unlabeled}) enforces the features extracted from noisy and clean radiographs to be consistent with each other, with a hyperparameter $\lambda_{\text{SSLR}}$ controlling the weighting between the two terms. The loss function is given by:
\begin{equation}
    \underset{\bm{\theta}_1,\,\bm{\theta}_2}{\min} \,\,  \mathbb{E}_{(\bm{\rho},\,\mathbf{D},\,\mathbf{T})} \frac{\|D_{\bm{\theta}_2}(E_{\bm{\theta}_1}(\mathbf{T}))- \bm{\rho}\|_2}{\| \bm{\rho}\|_2} + \lambda_{\text{SSLR}} \frac{\|E_{\bm{\theta}_1}(\mathbf{T}) - E_{\bm{\theta}_1}(\mathbf{D})\|_1}{ \| E_{\bm{\theta}_1}(\mathbf{D})\|_1}.
    \label{eq:unlabeled}
\end{equation}

\changes{Note that in the SSLR approach, the latent representation (output of the encoder), i.e., the features, are not directly supervised. The constraint on the output of the encoder, i.e., the latent space features, is that the features learned from the noisy radiograph should be consistent with the features learned from the clean radiograph. In other words, we are not supervising the latent space features to be a set of features that have been previously manually labeled, and the supervision is “implicit”. This is the reason why we call this variant a self-supervised latent representation~\cite{balestriero2023cookbook}.}

\subsection {ULR Approach}
Finally, we will also compare our feature extraction-based approaches to a network that assumes no prior on the learned features and reconstructs densities end-to-end using the encoder-decoder framework. We call this approach the unsupervised latent representation (ULR) approach, which is given as follows:

\begin{equation}
    \underset{\bm{\theta}_1,\bm{\theta}_2}{\min} \,\,  \mathbb{E}_{(\bm{\rho},\,\mathbf{T})} \frac{\|D_{\bm{\theta}_2}(E_{\bm{\theta}_1}(\mathbf{T})) - \bm{\rho}\|_2}{\|\bm{\rho}\|_2}
    \label{eq:no_reg}
\end{equation}


\section{\changes{Numerical Experiments}} \label{sec:experiments}
\changes{This section begins by outlining the numerical setup in Section~\ref{sec:exp_set}. The primary results and comparisons with baseline methods are then presented in Section~\ref{sec:main_results}. In Section~\ref{sec:reg_param_effect}, we investigate the influence of the regularization parameters $\lambda_{\text{PISLR}}$ and $\lambda_{\text{SSLR}}$ on both the reconstructed density accuracy and the extracted features. The subsequent sections examine the dependence of the reconstructed density on different types of noise and their scaling. Specifically, Section~\ref{sec:scatter_effect} analyzes the impact of coherent scatter and the order of background scatter on the reconstruction quality. Section~\ref{sec:gamma_photon} evaluates reconstruction accuracy under varying levels of gamma and photon noise. Collectively, these studies aim to assess the robustness of the proposed feature-based density estimation frameworks.}

\subsection{Numerical Setup}\label{sec:exp_set}
\begin{figure}[ht]
    \centering
    \includegraphics[width=0.9\linewidth]{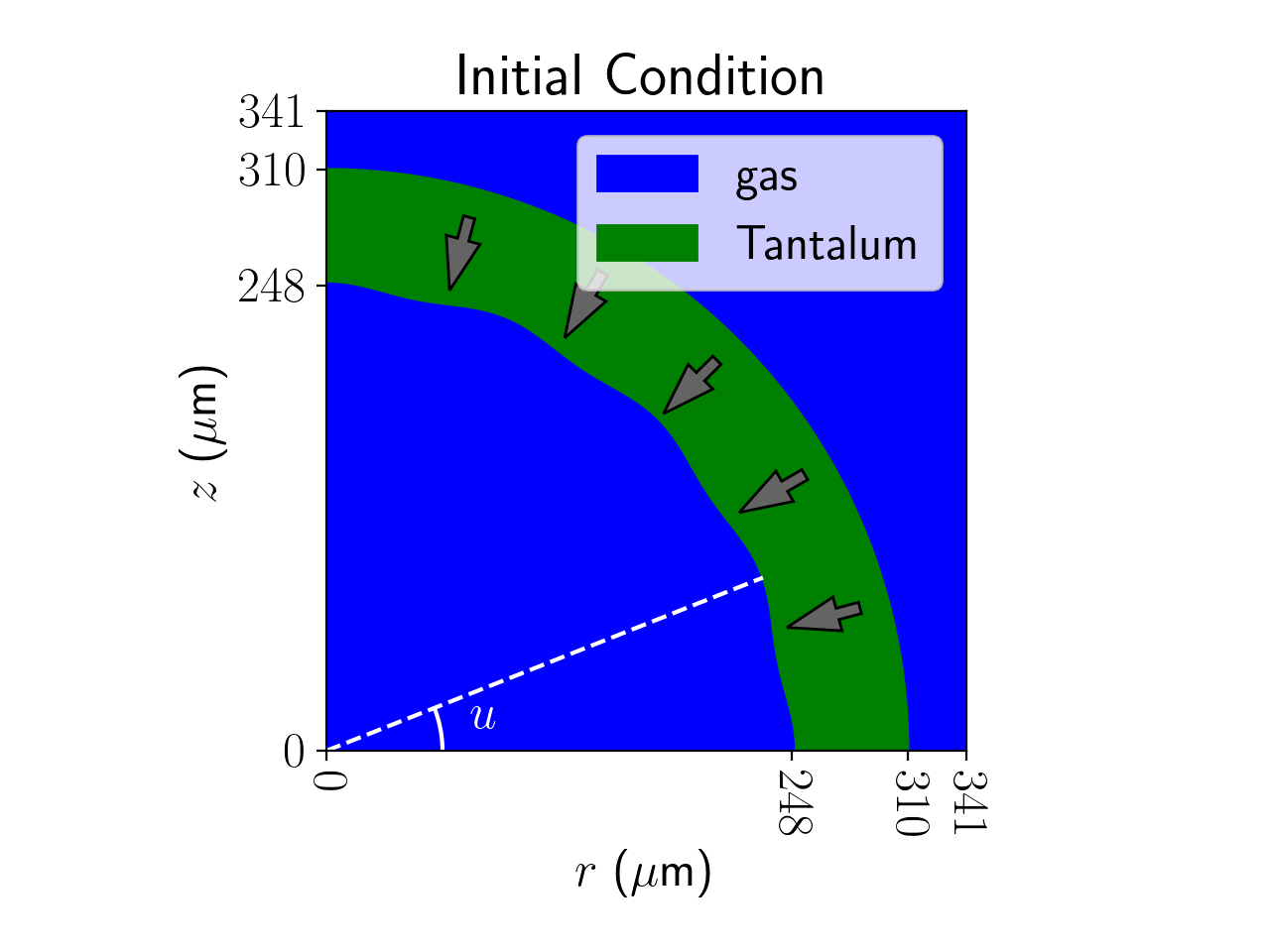}
    \caption{Representative ICF capsule for investigations.}
    \label{fig:ICF_capsule}
\end{figure}

For our numerical investigations, we simulate an ICF-double shell in which a Tantalum shell is driven into a gas.  The implosion of the shell creates a shock in the gas (air) that subsequently produces an outgoing shock which interacts with a perturbed metal (Tantalum) interface, creating a Richtmyer Meshkoff instability (RMI). A representative double shell is presented in Figure \ref{fig:ICF_capsule}. To further simplify the problem, we examine the implosion of a single e shell {made of} Tantalum as this configuration enables the salient features to be captured in the density field i.e., a complex gas-metal interface without the necessity to increase the complexity of the simulation. 

For this test problem, a dataset of simulations is generated using the parameters characterizing the Mie-Gr\"uneisen (MG) equation of state~\cite{robinson2019mie} given as
\begin{align}
    p \left( \chi:=1-\frac{\rho_0}{\rho}, T \right) 
    = \frac{\rho_0 c_s^2\chi\left(1 - \frac12\Gamma_0\chi\right)}
          {(1 - s_1\chi)^2}
    + \Gamma_0\rho_0 c_V (T - T_0),
\end{align}
where $\rho_0$ and $T_0$ are the reference density and temperature, respectively, $c_s$ is the speed of sound, $\Gamma_0$ is the Gr\"uneisen parameter at the reference state, $s_1$ is the slope of the linear shock Hugoniot, and $c_V$ is the specific heat capacity at constant volume.
The reference density $\rho_0$ and the reference temperature $T_0$ are kept fixed and the rest of the parameters $\{c_s, s_1, \Gamma_0, c_V\}$   are varied as shown in Table~\ref{tab:eos_param}.

\begin{table}[htbp]
  \centering
  \begin{tabular}{cccccc} 
  \hline \hline
    Profiles  & 1   & 2  & 3 & & \\
  \hline
    $\Gamma_0$  & 1.6   & 1.7  & 1.76 & 1.568 & 1.472 \\
  \hline
    $s_1$  & 1.32   & 1.464  & 1.342 &  & \\
  \hline
    $c_s$ [m/s]  & 339000   & 372900  & 305100 & 355000 &  \\
  \hline
    $c_V$ $[{\rm erg}\;{\rm g}^{-1}\;{\rm eV}^{-1}]$  & $1.6\times 10^{10}$   & $1.76 \times 10^{10}$  & $1.44 \times 10^{10}$ &    &  \\
  \hline
  \hline
  \end{tabular}
  \caption{Matrix of parameter values used to develop the simulated dataset. All combinations of the above parameters are used to simulate our data.}
  \label{tab:eos_param}
\end{table}

\begin{figure}[ht]
    \centering
    \includegraphics[width=\linewidth]{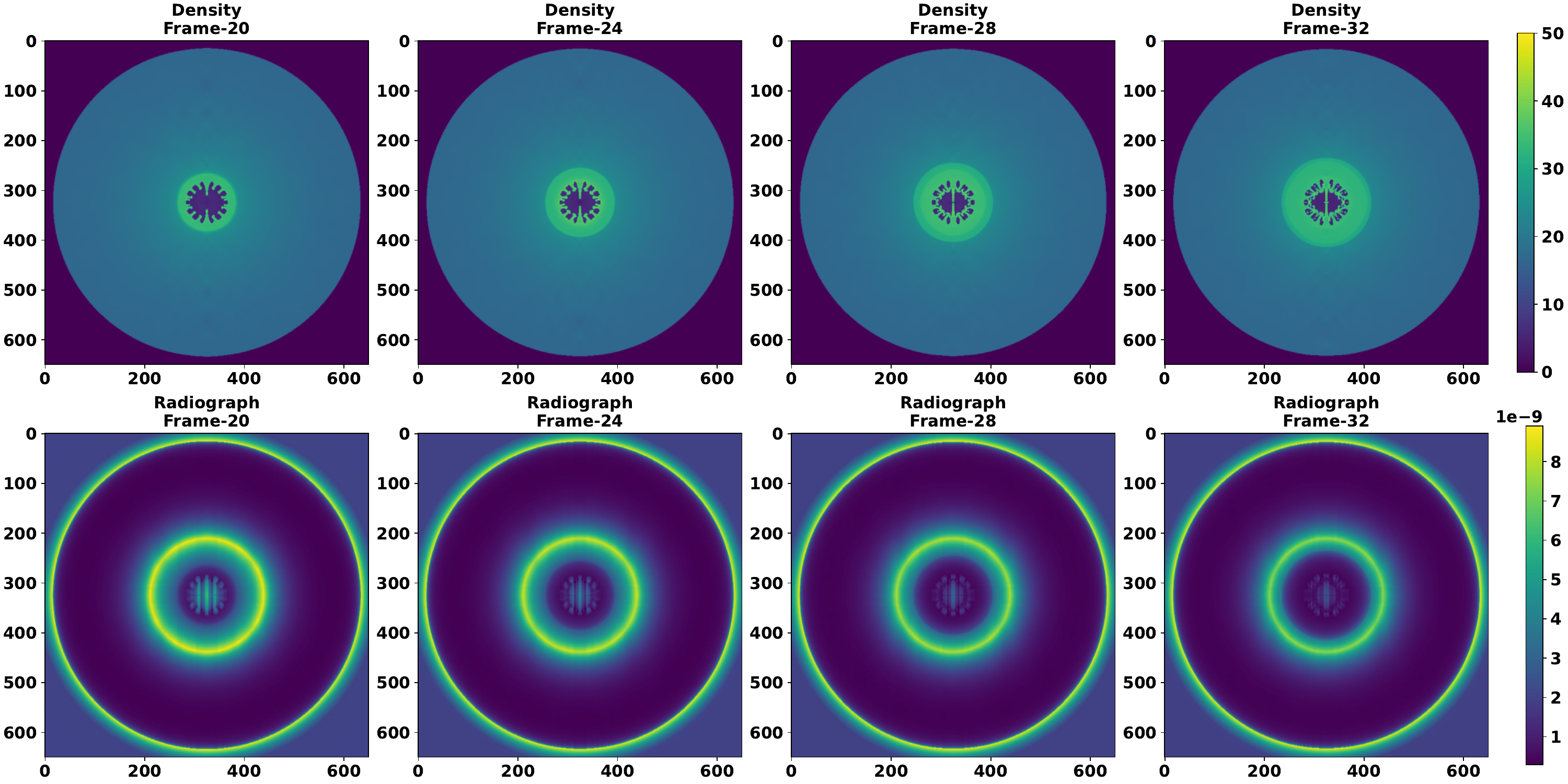}
    \caption{2D profiles of a density sequence and the corresponding synthetic clean radiographs.}
    \label{fig:den_rad}
\end{figure}

\begin{figure}[ht]
    \centering
    \includegraphics[width=0.9\linewidth]{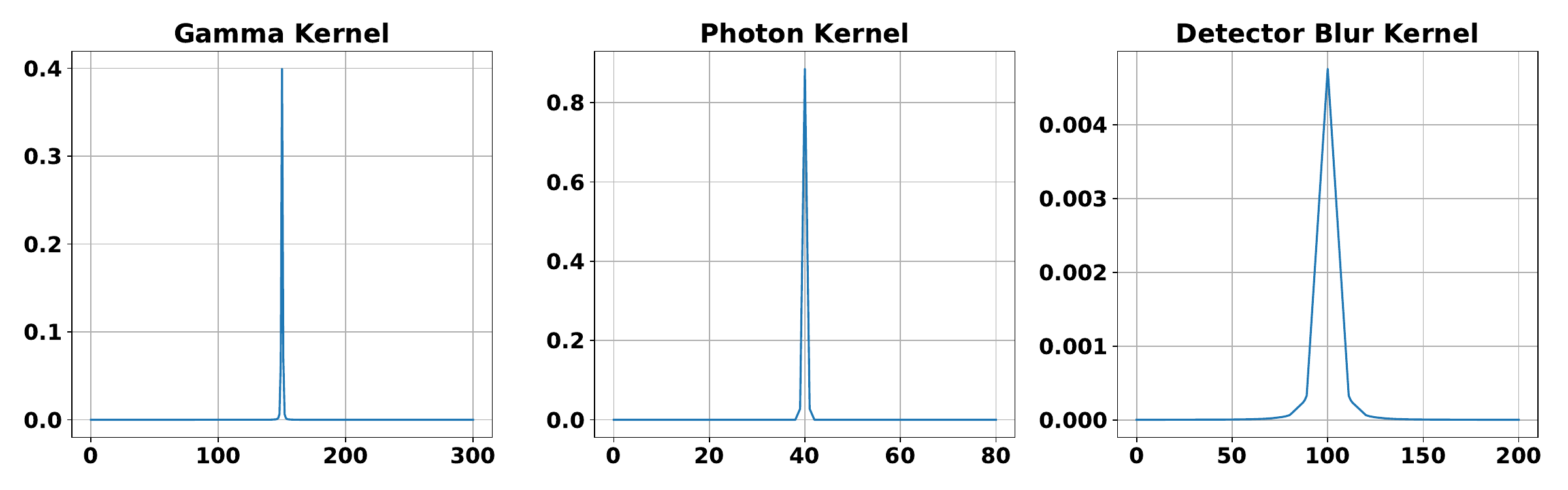}
    \caption{{1D lineout of the gamma, photon, and detector blur kernels used for generating various kinds of perturbations.}}
    \label{fig:kernels}
\end{figure}

The hydrodynamic dataset consists of 10074 simulations consisting of different combinations of the physical parameters. Each simulation file contains the dynamics for 41 time frames with each density profile being a $440\times 440$ quadrant of the full profile. For our numerical investigations, we randomly chose 9067 such simulations for training and 1007 for testing. For each sequence, we choose 4 equally spaced time frames chosen from each sequence - the $20^{th}$, $24^{th}$, $28^{th}$ and $32^{nd}$ frames out of $41$ that capture an important part of the dynamics. For simulating radiographs and network training, we generate a full 2D density profile by flipping left-right and top-bottom the given one quadrant and cropping it to the central $650\times 650$ region that contains the full object. The 2D profile is taken as a slice of an axis-symmetric 3D object. 
We used 3D cone beam CT geometry in the ASTRA toolbox~\cite{van2015astra, van2016fast} to simulate the areal densities (and subsequently the direct radiograph) from the clean 3D density profiles (see example in Fig.~\ref{fig:den_rad}). 
Scatter and noise are then incorporated into the model using the models discussed in Section~\ref{sec:dynamic_radiograpy}. The parameter ranges used in the blur, scatter, and noise models for obtaining the noisy radiographs are provided in Table~\ref{tab:noise_params}. These parameters, referred to as in-population parameters, 
were used during the training of various models. At testing time, we also vary parameters outside these ranges (referred to as out-of-population setup) to investigate the robustness of various models.
The kernels $\bm{\phi}_{db}$, $\bm{\phi}_{g}$ and $\bm{\phi}_{p}$ were pre-generated based on the setup {and are shown in Figure~\ref{fig:kernels}}.

\begin{table}[htbp]
  \centering
  \begin{tabular}{cccccccc} 
  \hline \hline
    Parameter  & $\sigma_{blur}$ & $\sigma_{scatter}$ & $\kappa$ & $a,b$  & $n$ &  $\kappa_g$ &   $\kappa_p$ \\
  \hline
    Value  & $[1,1.3]$ & $[10,30]$ & $[0.1,0.3]$ & $[-3.9, 3.9]\times 10^{-5}$ & $1$ & $1$ & $1$
  \end{tabular}
  \caption{Ranges of parameters used to simulate noisy radiographs during training. For those parameters with specified ranges, a random value in the range was selected for every simulation.}
  \label{tab:noise_params}
\end{table}

We used a single-channel four-layer U-Net \cite{ronneberger2015u} architecture for both the encoder and decoder networks in our frameworks and comparisons. The networks map the noisy and corrupted 2D radiographs (of axis-symmetric 3D objects) to a 2D slice of the density profiles (reconstructions). The corresponding loss functions were optimized using the Adam optimizer~\cite{kingma2014adam} with a learning rate of $10^{-3}$. The networks were trained for 100 epochs on an Nvidia RTX A5000 GPU with 24 GB RAM. Each time the network selects a batch size of 4 which corresponds to the four selected frames of a particular sequence. We update the parameters of the encoder and decoder network in an alternating manner after every $3$ iterations. While testing, the metric used to evaluate the accuracy of reconstructed density profiles is the root mean squared error (RMSE) given as
\begin{equation}
\text{RMSE} = \sqrt{\frac{1}{N} \sum_{i=1}^N (\bm{\rho}_{clean_i}-\bm{\rho}_{reconstructed_i})^2}
\label{eq:rmse}
\end{equation}
where $\bm{\rho}_{clean_i}$ and $\bm{\rho}_{reconstructed_i}$ are the i$^{th}$ pixel values of the clean and reconstructed densities (2D slices of 3D profiles), and $N$ is their number of pixels.

\subsection{Main Results and Comparisons}\label{sec:main_results}
\begin{figure}[ht]
    \centering    
    \includegraphics[width=0.9\linewidth]{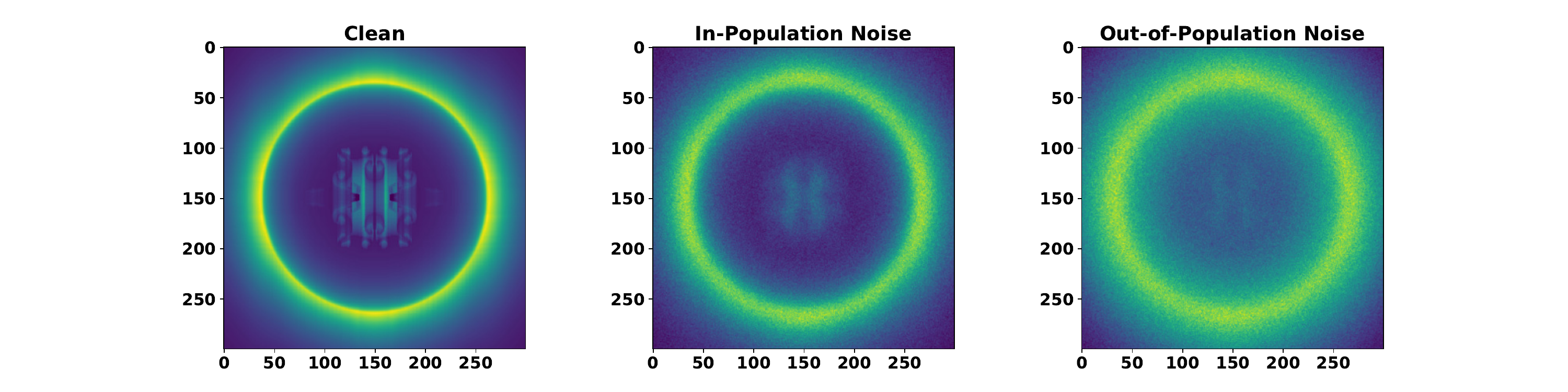}
    \caption{2D profiles of clean and noisy synthetic radiographs (in and out of population) over the central $300 \times 300$ region including the shock. The parameters used for in-population noise are $\sigma_{\text{scatter}}=10$, $\kappa=0.2$, order $=1$ while the out-of-population parameters are $\sigma_{\text{scatter}}=40$, $\kappa=3$, order$=2$.}
    \label{fig:noisy_rad_vs_scaling}
\end{figure}

\begin{figure}[ht]
    \centering
    \includegraphics[width=0.7\linewidth]{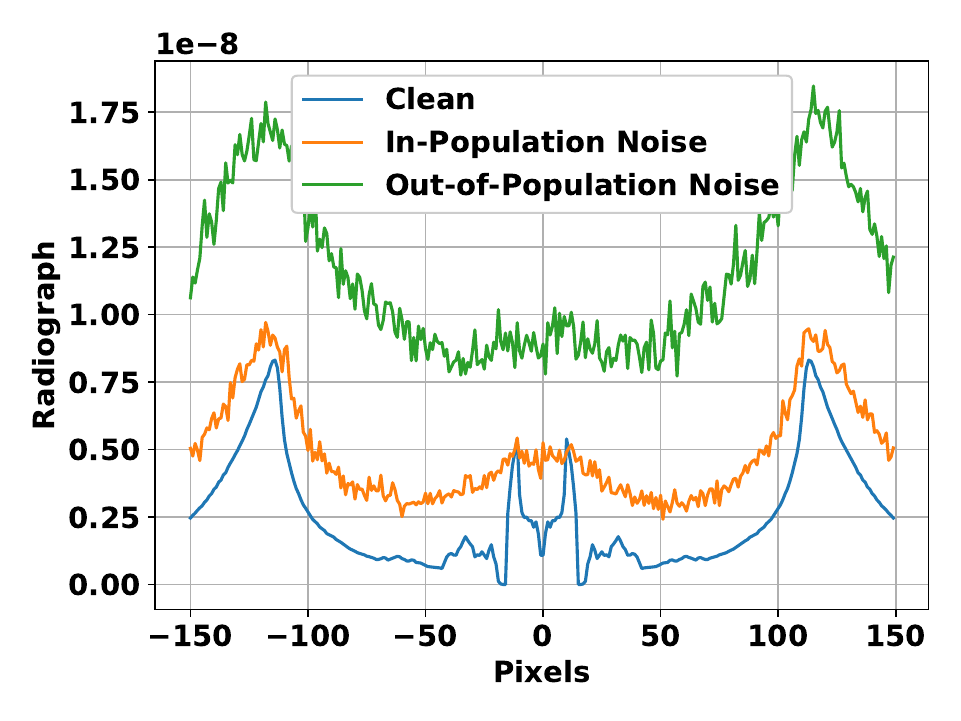}

    \caption{1D lineouts of clean and noisy radiographs (in and out of population).  In-population parameters: $\sigma_{\text{scatter}}=10$, $\kappa=0.2$, order $=1$. Out-of-population parameters: $\sigma_{\text{scatter}}=40$, $\kappa=3$, order$=2$. Only the central $300 \times 300$ region is shown.}
    \label{fig:noisy_rad_lineout}
\end{figure}

In this section, we show a comparison of the performance and robustness of our feature-constrained ({PISLR and SSLR) density reconstruction approaches to the ULR approach mentioned} in Eq.~\eqref{eq:no_reg}. \changes{We also compare our approaches with a GAN-based approach for density reconstruction~\cite{huang2022physics} from the radiographs corrupted by the same noise and scatter. The official implementation was adopted from the authors' code repository.~\footnote{\url{https://github.com/zhishenhuang/hydro/blob/main/wgan_train.py}} The training loss was defined as a weighted sum of the GAN loss and the supervised density reconstruction loss. The network was trained for 100 epochs with a batch size of 3. Both the generator and discriminator were optimized using the Adam optimizer, with initial learning rates of $10^{-5}$ and $2 \times 10^{-5}$, respectively. To enhance generalization, a StepLR scheduler was applied based on the validation loss. The initial weight assigned to the density reconstruction loss term in the generator’s loss function was set to 0.99 and was decayed by a factor of 0.97 each epoch.} 
The testing is first performed for in-population scatter and noise levels, where the properties of scatter and noise are within the regime used for training the models. {Subsequently, to explore the robustness of each reconstruction approach, we test our models on noisy radiographs generated using out-of-population noise levels where the noise parameters used were outside the training regime}. The parameters used for {generating} in-population noise for all the studies are $\sigma_{\text{scatter}}=10$, $\kappa=0.2$, order $=1$ while the out-of-population parameters utilized are $\sigma_{\text{scatter}}=40$, $\kappa=3$,  and order$=2$. The additional parameters were randomly selected from the fixed range prescribed in Table~\ref{tab:noise_params}. Figures~\ref{fig:noisy_rad_vs_scaling} and~\ref{fig:noisy_rad_lineout} depict the 2D profiles and 1D lineouts of the clean and noisy synthetic radiographs for both in-population as well as out-of-population noise.

\begin{figure}[ht]
    \centering
    \includegraphics[width=\textwidth]{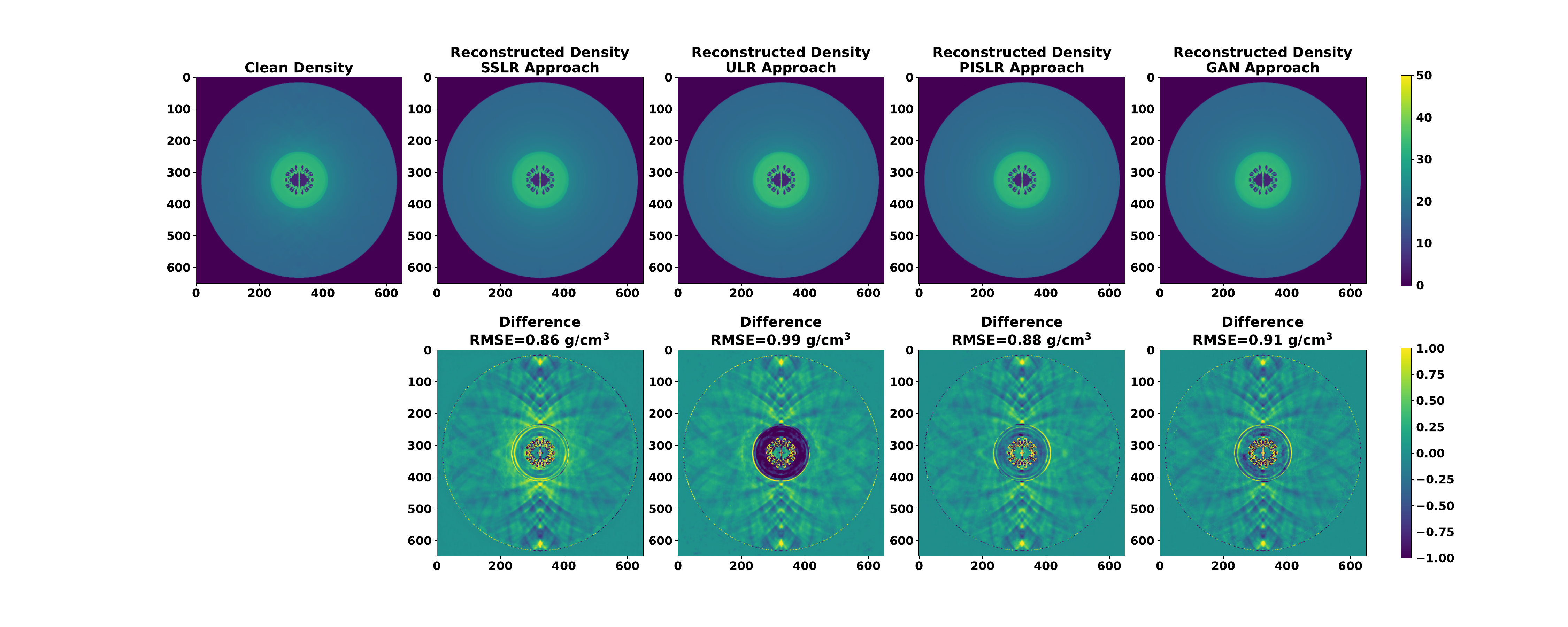}
\caption{2D profiles of the $32^{nd}$ frame for clean and reconstructed densities along with the difference (error) images with different methods for in-population noise level. The SSLR approach gives the best-reconstructed density among all four compared approaches with the lowest RMSE.}
    \label{fig:rho_recon2d}
\end{figure}

\begin{figure}[ht]
    \centering
    \includegraphics[width=\textwidth]{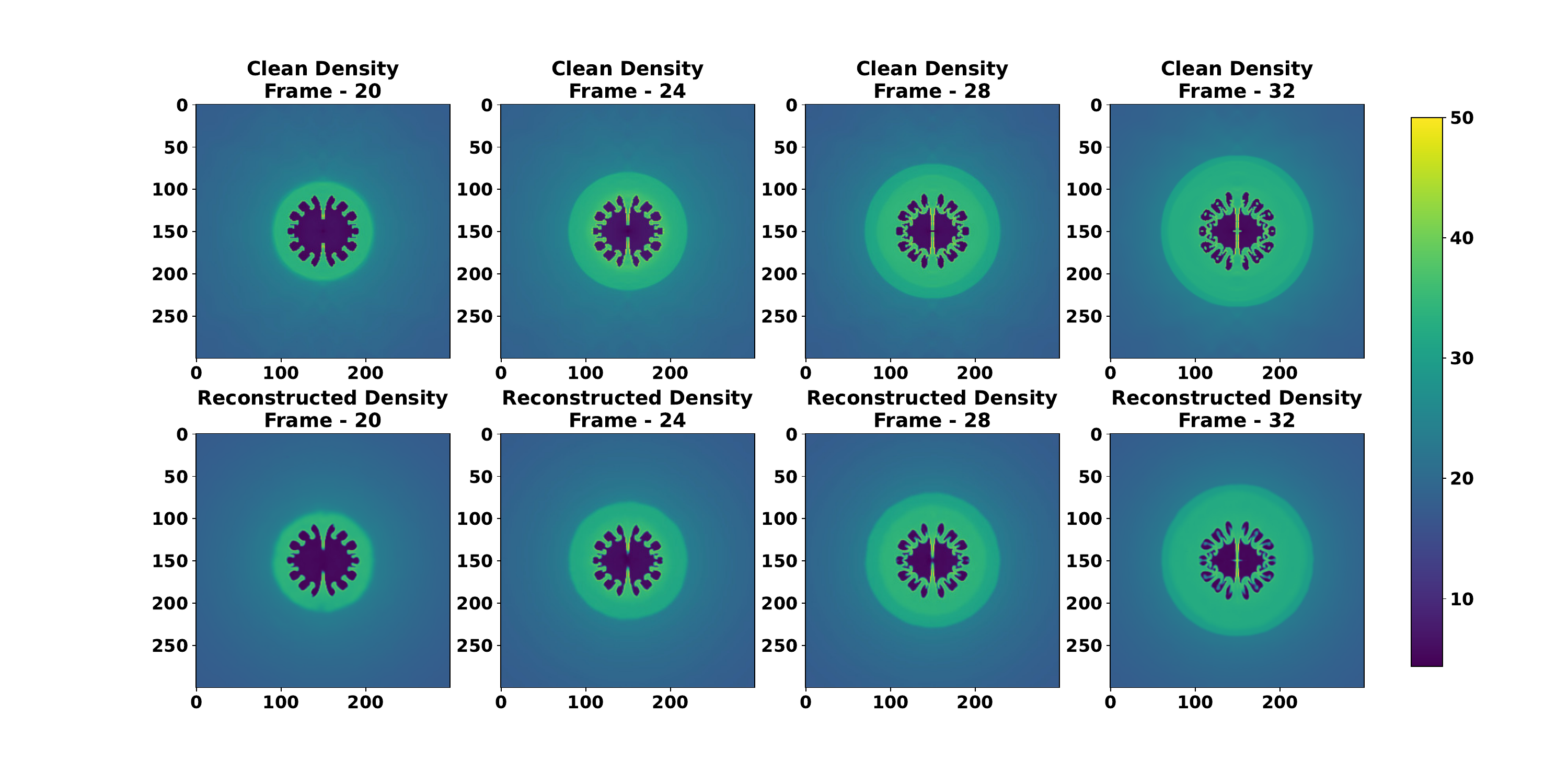}
\caption{2D profiles for clean and reconstructed densities using the SSLR approach for four frames together over the central $300 \times 300$ region for in-population noise parameters. The reconstructed density accurately captures the outgoing shock.}
    \label{fig:rho_recon2d_all_frames}
\end{figure}

\begin{figure}[ht]
    \centering
    \includegraphics[width=\textwidth]{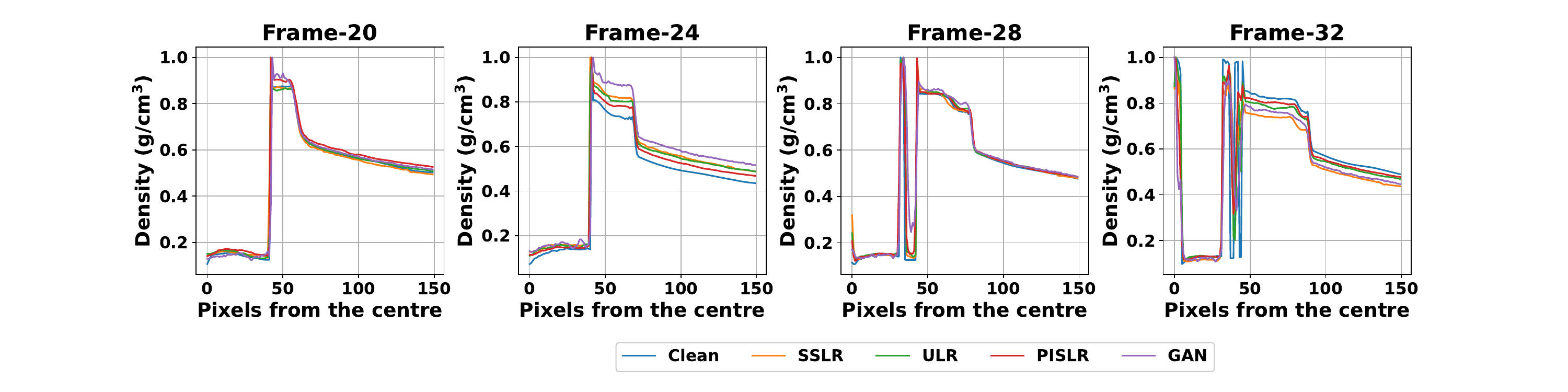}
    \caption{1D lineout for clean and reconstructed densities for three approaches restricted to the region of interest. Only the first 150 pixels from the center are shown.}
    \label{fig:rho_recon1d}
\end{figure}

Figure~\ref{fig:rho_recon2d} presents the 2D profiles of reconstructed and clean densities along with the difference between them (cut-off to the range of $[-1,\,1]$) for the different reconstruction approaches. {The RMSE for the reconstructed densities is provided on top of each frame and is in the range $0.86-0.99$ g/cm$^3$}. From the figure, we can observe that the best reconstruction accuracy, along with accurate localization of the gas-metal interface, is achieved with the SSLR approach. The evolution of the clean and reconstructed densities (using the SSLR approach) over different frames is captured in Figure~\ref{fig:rho_recon2d_all_frames}. The corresponding 1D lineouts for the clean and denoised densities are presented in Figure~\ref{fig:rho_recon1d}. {From these figures, we can see that the reconstructed density captures the outgoing shock accurately.} 

\begin{figure}[ht]
    \centering
    \begin{subfigure}[t]{0.5\textwidth}
        \centering
        \includegraphics[width=0.9\linewidth]{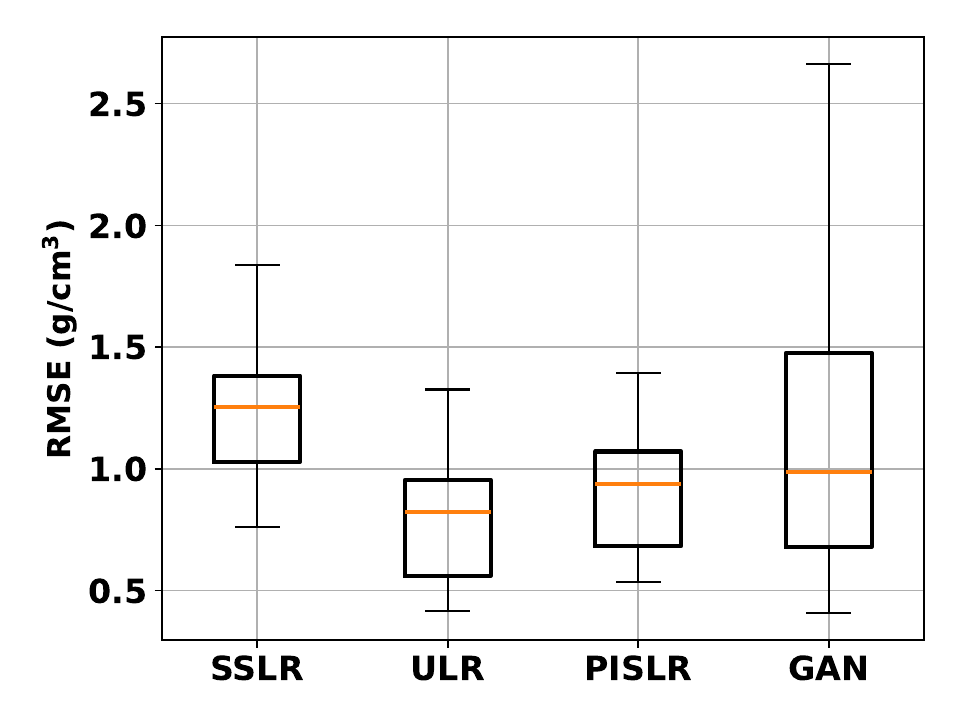}
        \caption{In-population noise}
    \end{subfigure}%
    \begin{subfigure}[t]{0.5\textwidth}
        \centering
        \includegraphics[width=0.9\linewidth]{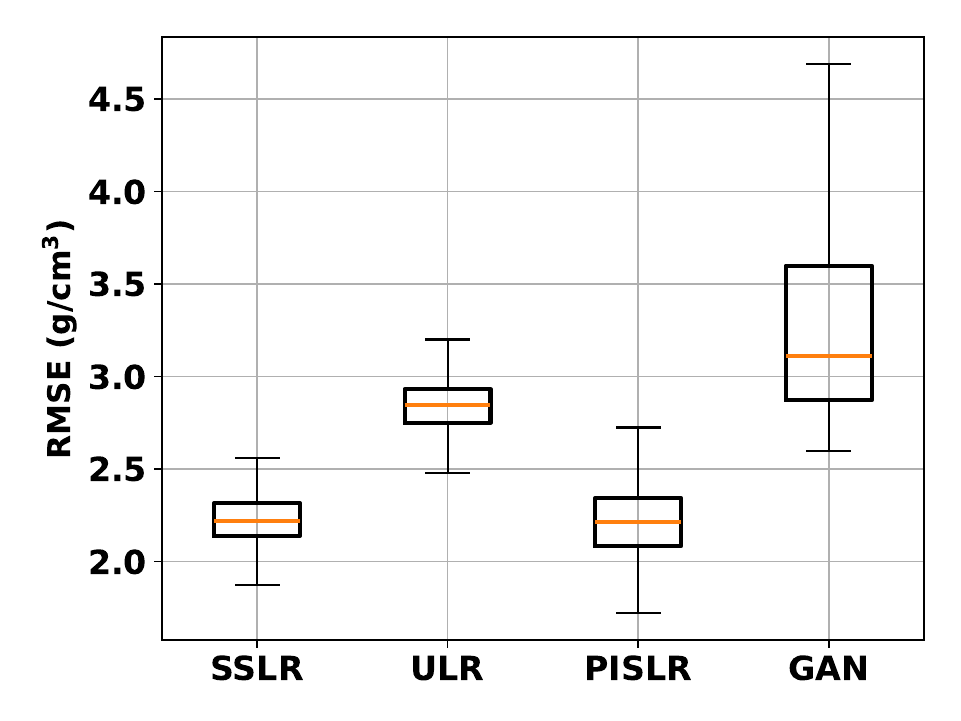}
        \caption{Out of population noise}
    \end{subfigure}
    \caption{\changes{Boxplot of RMSE (in g/cm$^3$) values of reconstructed densities over 1007 test cases for a) in-population and b) out-of-population noise levels. Approaches used: 1) SSLR, 2) ULR, 3) PISLR, and 4) GAN.}}
    \label{fig:rmse_boxplots}
\end{figure}

{The reconstruction results over all the $1007$ test cases are summarized using boxplots shown in Figure~\ref{fig:rmse_boxplots}. The minimum and maximum values are indicated by whiskers and the boxes indicate the first and third quartiles. The median value of the distribution is indicated by the orange line at the center.} \changes{From the figure, it can be observed that all approaches give somewhat similar accuracy for in-population noise levels. The out-of-population boxplots indicate that the SSLR and PISLR features are the most robust to unknown scatter/noise.}

\begin{figure}[ht]
    \centering
    \includegraphics[width=\linewidth]{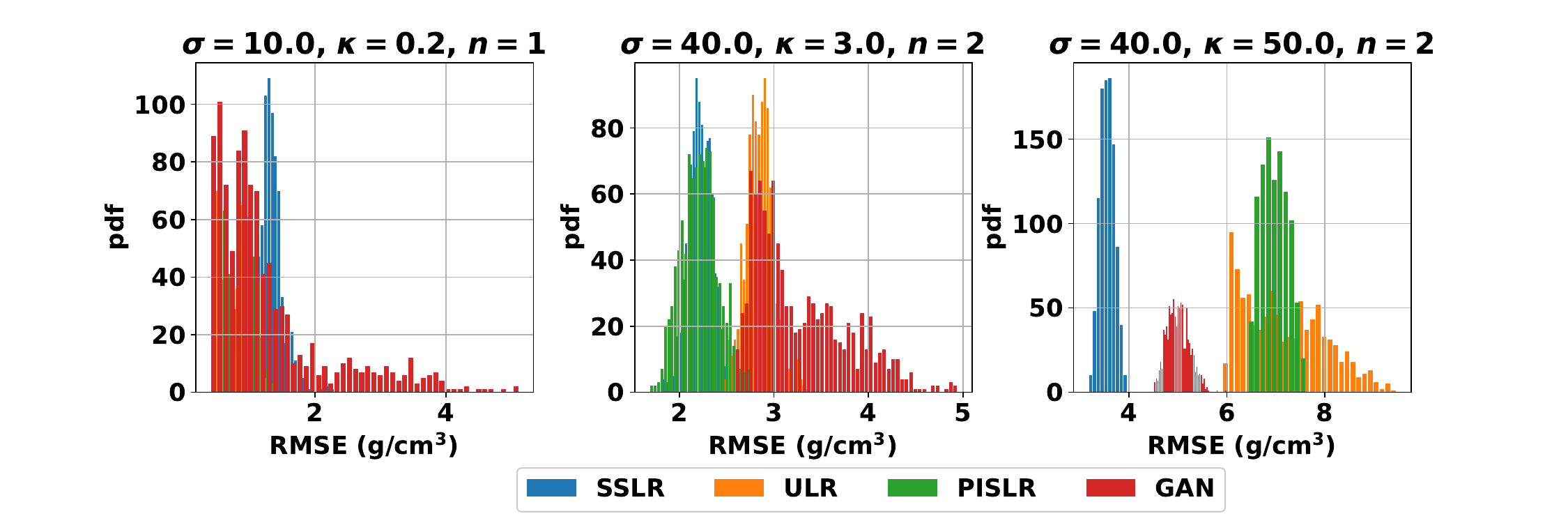}
    \caption{\changes{Histogram of RMSE (in g/cm$^3$) of reconstructed densities for in and out of population noise levels. Approaches used: 1) SSLR (blue), 2) ULR (orange), 3) PISLR (green), and 4) GAN (red).}}\label{fig:rmse_histogram}
\end{figure}

\changes{We also summarize the reconstruction performance using histograms of the RMSE values for each method, as shown in Figure~\ref{fig:rmse_histogram}. Results are presented separately for in-population and out-of-population noise levels, with the latter evaluated at two different scaling factors: $\kappa = 3$ (moderate noise) and $\kappa = 50$ (high noise).
For in-population noise, the RMSE distributions show that all four methods—ULR, PISLR, GAN, and SSLR—perform comparably, with ULR, PISLR, and GAN achieving the lowest reconstruction errors and SSLR following closely with only slightly higher RMSE values. However, under out-of-population noise with $\kappa = 3$, the performance of ULR and GAN degrades noticeably, as indicated by a rightward shift in their RMSE distributions. In contrast, the feature-based PISLR and SSLR methods demonstrate greater robustness, maintaining lower reconstruction errors in this regime.
At the highest noise level ($\kappa = 50$), the SSLR approach remains the most robust, with its RMSE distribution centered around 3.5 g/cm$^3$, while the errors for ULR and GAN increase substantially, with their distributions centered around 7 g/cm$^3$.}

\subsection{Effect of Hyperparameters on the Feature-based Approaches}\label{sec:reg_param_effect}
In this subsection, we present the effect of changing the hyperparameters $\lambda_{\text{PISLR}}$ and $\lambda_{\text{SSLR}}$ on the observed features and the corresponding accuracy of reconstructed densities.

\subsubsection{Visualization of Features}
\begin{figure}[ht]
    \centering
    \includegraphics[width=0.9\linewidth]{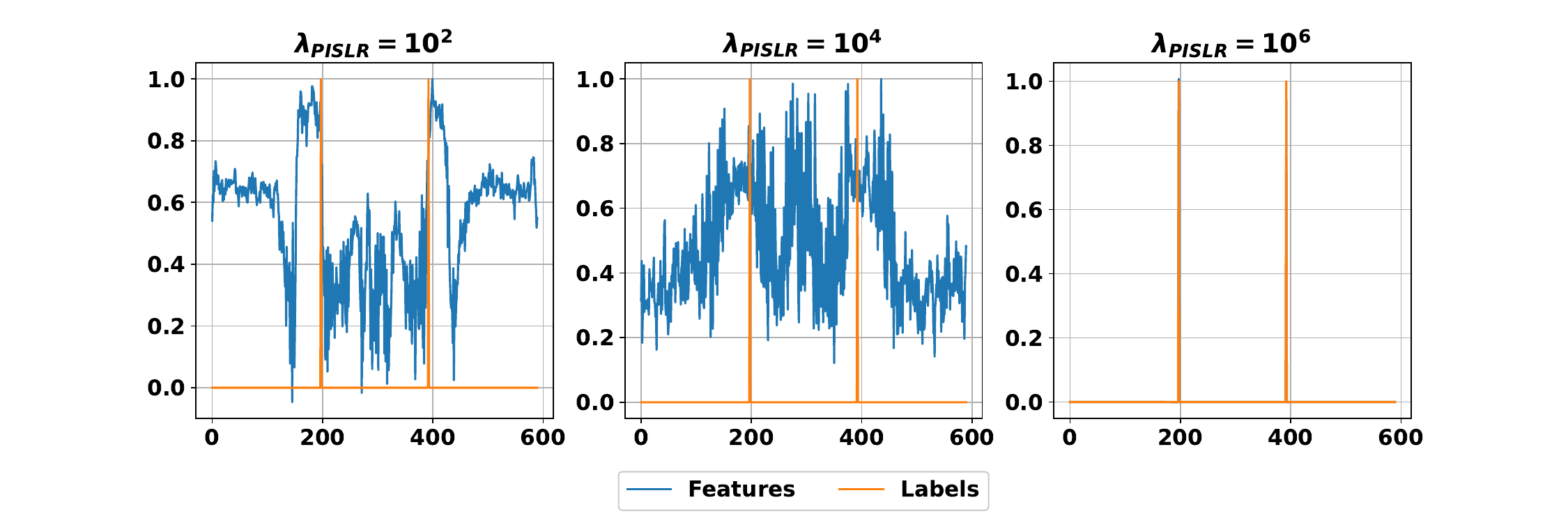}
    \caption{Features for one frame from encoder trained using the shock edgemap as the labels using $\ell_1$ loss. The subplots shown are for networks trained with $\lambda_{\text{PISLR}}=10^2$, $\lambda_{\text{PISLR}}=10^4$ and $\lambda_{\text{PISLR}}=10^6$. The network trained with $\lambda_{\text{PISLR}}=10^6$ gives the features closest to the provided labels. Features are normalized to $[0,1]$.}
    \label{fig:masked_features_all_lamdas_l1_loss}
\end{figure}

\begin{figure}[ht]
    \centering
    \includegraphics[width=0.7\linewidth]{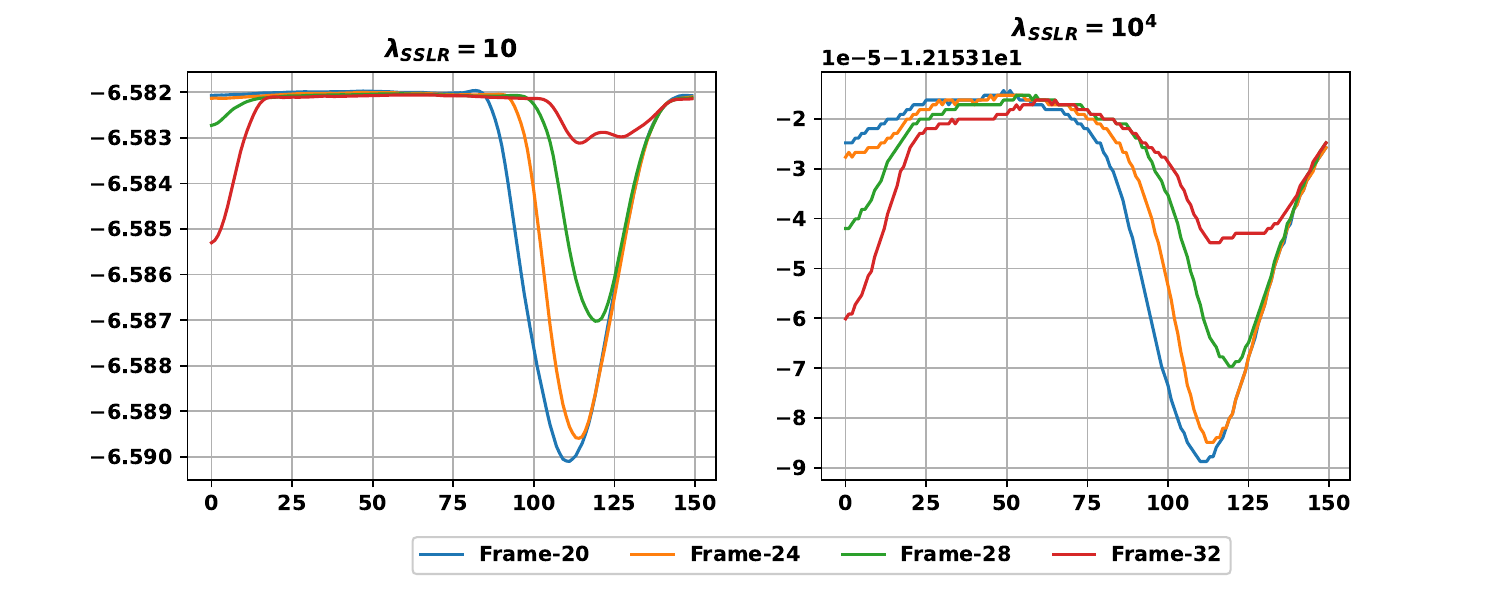}
    \caption{1D lineout of features from encoder trained using the SSLR approach for four different frames. The subplots shown are for networks trained with $\lambda_{\text{SSLR}}=10$ and $\lambda_{\text{SSLR}}=10^4$. Only the first 150 pixel values from the center are shown.}
    \label{fig:unlabeled_features_2_lamdas}
\end{figure}

To examine the sparsity in the features predicted by the PISLR features network, the networks in Eq.~\eqref{eq:masked} were trained using different values of the hyperparameter $\lambda_{\text{PISLR}}$.
Figure~\ref{fig:masked_features_all_lamdas_l1_loss} presents the features obtained with different choices of the hyperparameter $\lambda_{\text{PISLR}}$. From the figures, we can observe that the network trained with hyperparameter $\lambda_{\text{PISLR}} = 10^6$ exhibits features that best match the labels out of all four networks. Similarly, figure~\ref{fig:unlabeled_features_2_lamdas} shows the features produced by the encoder trained using the SSLR approach (loss function given by Eq.~\eqref{eq:unlabeled}) for different values of the hyperparameter $\lambda_{\text{SSLR}}$. The SSLR features are smoother and seem to evolve similar to the shock over time. We hypothesize that the smoothness of the features is responsible for the resiliency to noise in the synthetic radiographs.

\subsubsection{Impact of the Hyperparameter on Reconstruction Accuracy}
\begin{figure}[ht]
    \centering
    \begin{subfigure}[t]{0.5\textwidth}
        \centering
        \includegraphics[width=0.9\linewidth]{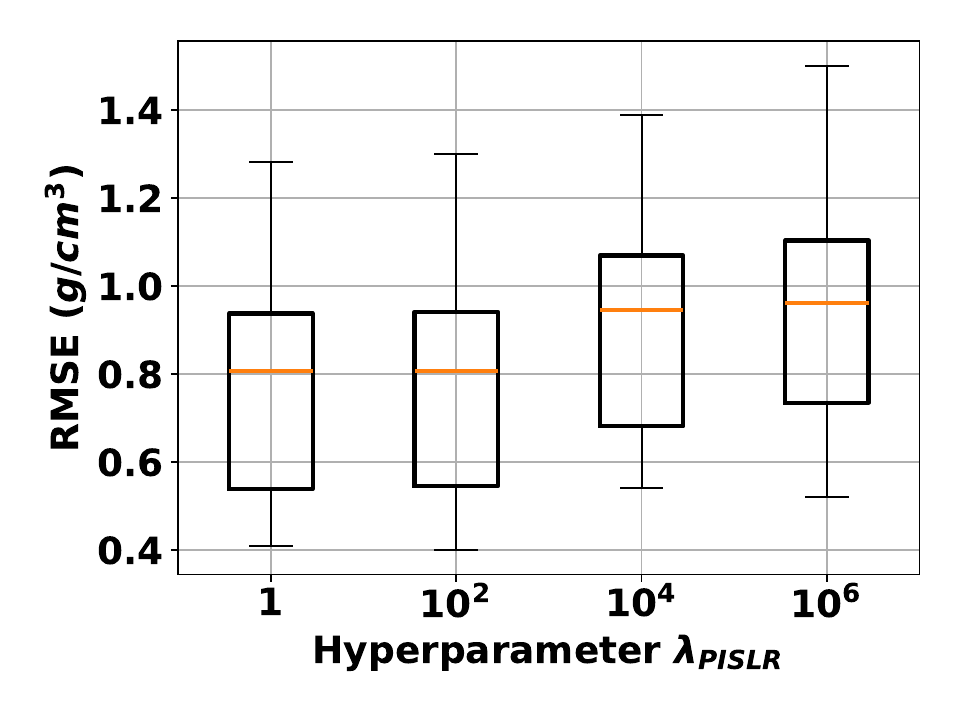}
        \caption{In-population noise}
    \end{subfigure}%
    \begin{subfigure}[t]{0.5\textwidth}
        \centering
        \includegraphics[width=0.9\linewidth]{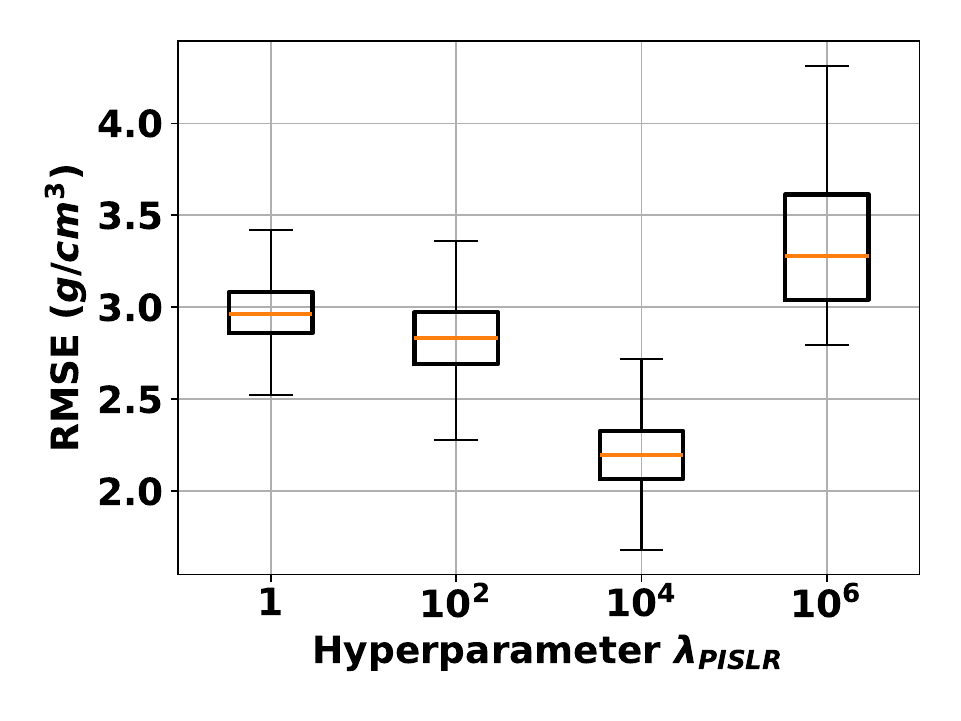}
        \caption{Out of population noise}
    \end{subfigure}
     \caption{Boxplots of RMSE for reconstructed densities over all test cases with the encoder trained using the shock edgemap (as labels) with $\ell_1$ regularizer loss using different values of hyperparameter $\lambda_{\text{PISLR}}$. The subplots shown are for in and out-of-population noise parameters.}  \label{fig:boxplots_masked_all_lambas_l1loss}
\end{figure}

\begin{figure}[t!]
    \centering
    \begin{subfigure}[t]{0.5\textwidth}
        \centering
        \includegraphics[width=0.9\linewidth]{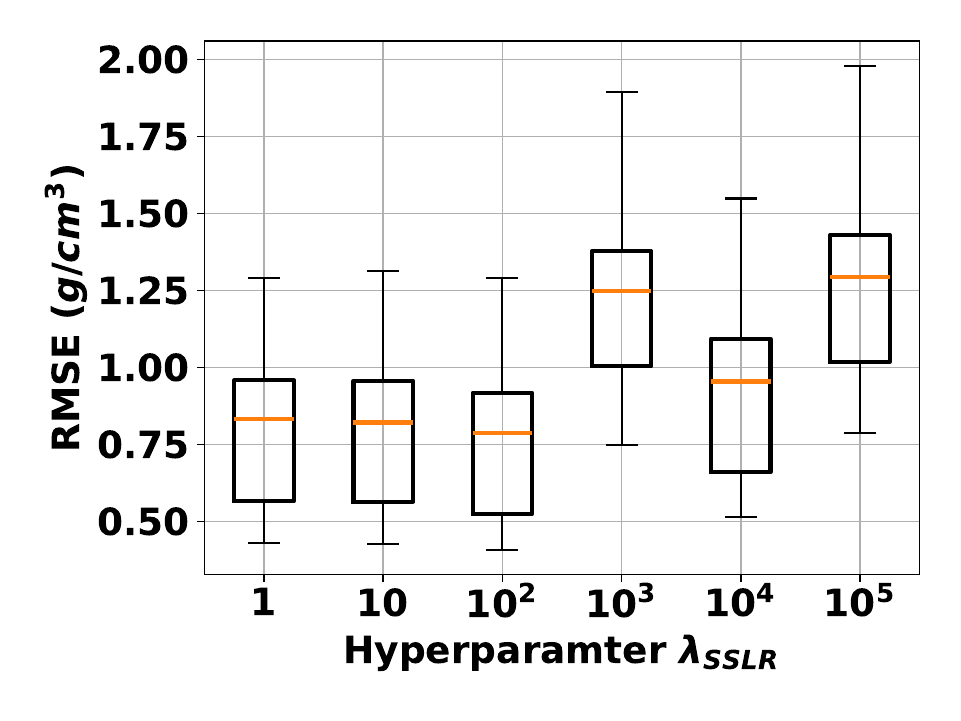}
        \caption{In-population noise}
    \end{subfigure}%
    \begin{subfigure}[t]{0.5\textwidth}
        \centering
        \includegraphics[width=0.9\linewidth]{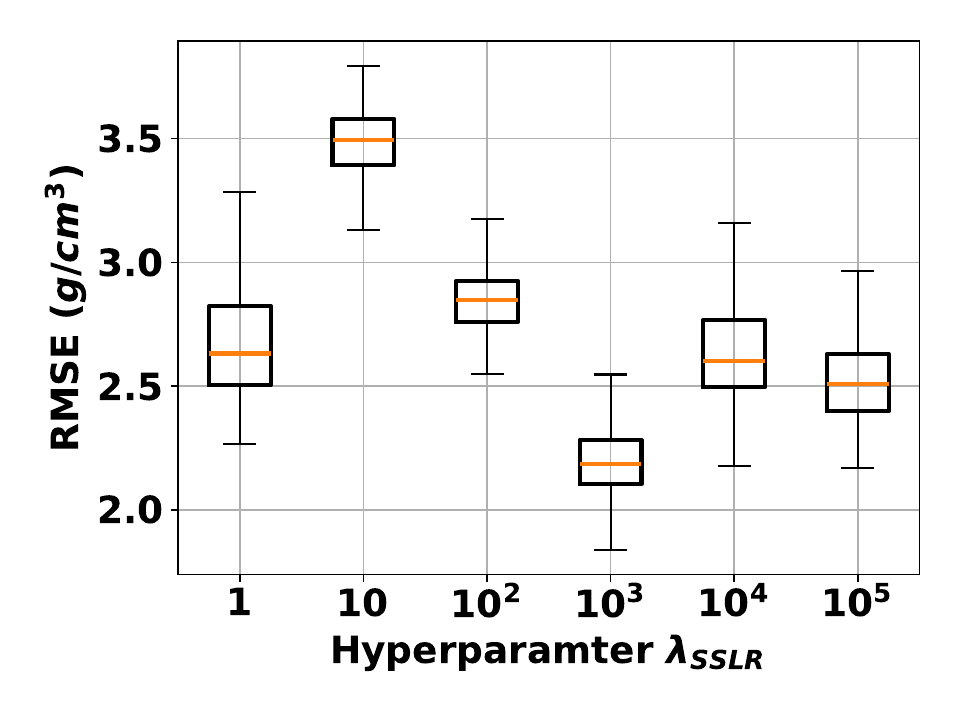}
        \caption{Out of population noise}
    \end{subfigure}
     \caption{Boxplot of RMSE for reconstructed densities over all test cases with the encoder-decoder trained for the SSLR approach using different values of hyperparameter $\lambda_{\text{SSLR}}$. The subplots shown are for in and out-of-population noise parameters.}  \label{fig:boxplots_robust_all_lambas}
\end{figure}

{To evaluate the effect of the hyperparameters $\lambda_{\text{PISLR}}$ and $\lambda_{\text{SSLR}}$ on the reconstruction quality, we noted down the RMSE of reconstructed densities over all the test sequences for different values of these hyperparameters.} Figure~\ref{fig:boxplots_masked_all_lambas_l1loss} presents such a boxplot for the PISLR approach (Eq.~\eqref{eq:masked}) for both in and out of population noise levels. Examination of Figure~\ref{fig:boxplots_masked_all_lambas_l1loss} indicates that the network trained with $\lambda_{\text{PISLR}}=10^4$ results in the best RMSE trade-offs for reconstructed densities. We use this trained model as our best model and compare it with the baselines in the subsequent studies. As an alternative for better capturing the features, we also explored binary cross entropy (BCE) loss instead of the $\ell_1$ loss as the training loss regularizer and found that the normalized $\ell_1$ regularizer gave better overall density reconstructions. In a similar {manner}, we repeated this study for the encoder-decoder trained for the SSLR approach. The RMSE variation (boxplots) over all the test cases for different values of hyperparameter $\lambda_{\text{SSLR}}$ in the loss function of Eq.~\eqref{eq:unlabeled} is shown in Figure~\ref{fig:boxplots_robust_all_lambas}. We observe that $\lambda_{\text{SSLR}}=10^3$ provides good RMSE trade-offs over different noise regimes, and we use it as the base model for the SSLR approach to compare it against baselines.

\subsection{Examination of Scatter Scaling and Background Scatter on Density Reconstructions}\label{sec:scatter_effect}
\begin{figure}
    \centering
    \begin{subfigure}[t]{0.45\textwidth}
        \centering
        \includegraphics[width=0.9\linewidth]{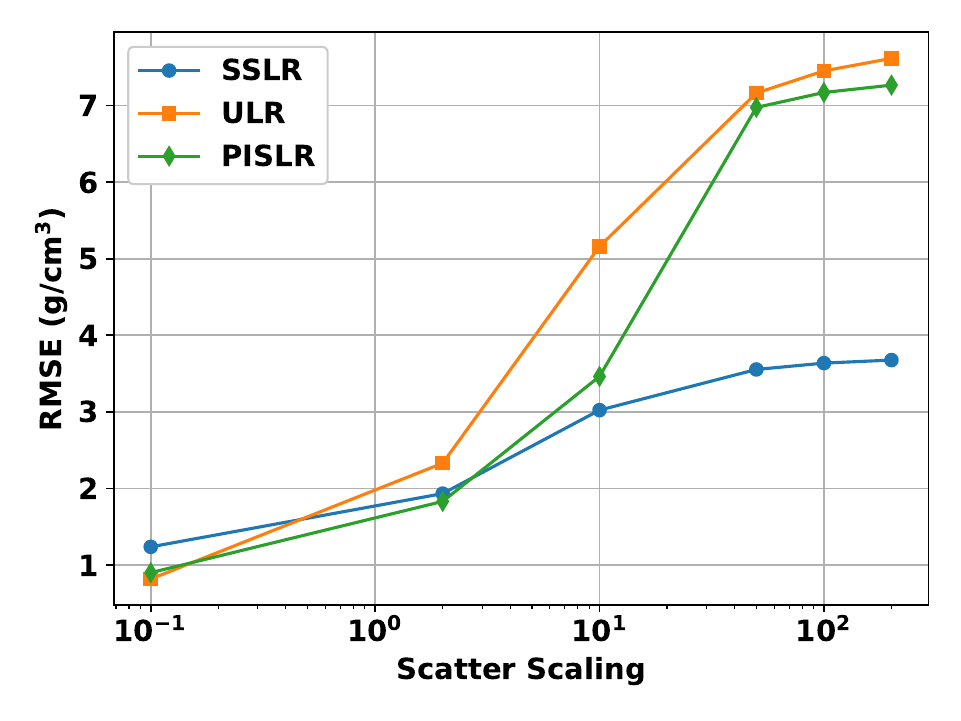}
        \caption{RMSE vs Scaling ($\sigma_{scatter}=40$, $n=2$)}
        \label{fig:rmse_vs_scaling}
    \end{subfigure}
    \begin{subfigure}[t]{0.45\textwidth}
        \centering
        \includegraphics[width=\linewidth]{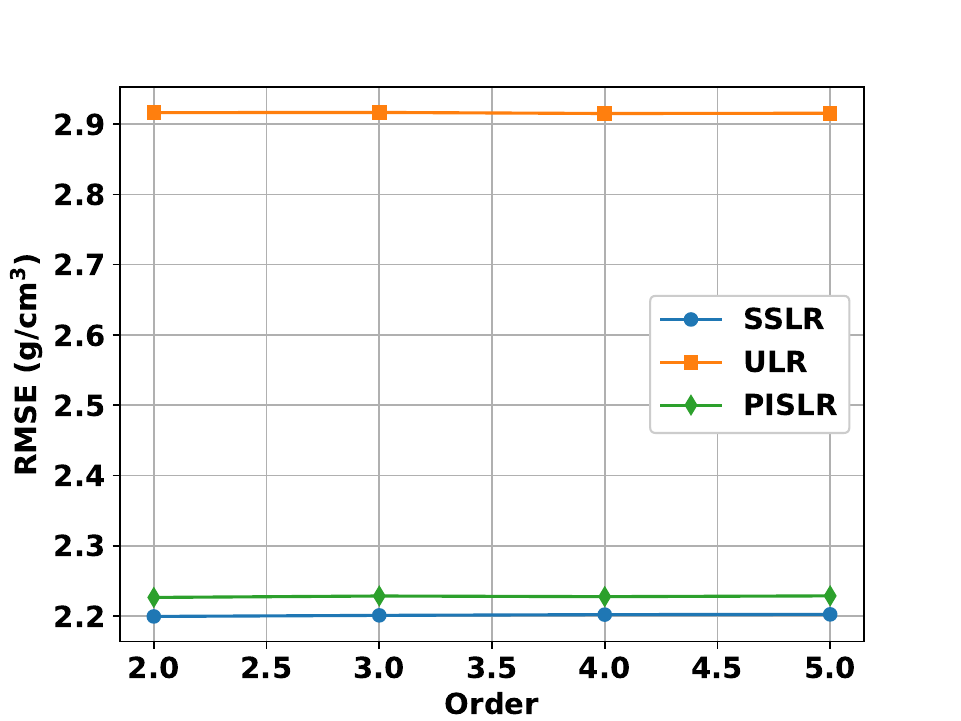}
        \caption{RMSE vs Order ($\sigma_{scatter}=40$, $\kappa=3$)}
        \label{fig:rmse_vs_order}
    \end{subfigure}
    \caption{Mean RMSE of reconstructed densities over 1007 test cases as a function of a) scaling of correlated scatter $\kappa$ and b) order of background scatter polynomial $n$.}
\end{figure}

In this section, we examine the effect of increasing the scatter scaling as well as the background scatter on the quality of reconstructed densities. Figure~\ref{fig:rmse_vs_scaling} presents the variation of mean RMSE over 1007 test cases for all approaches as a function of scatter scaling $\kappa$. For low scatter levels, the ULR approach performs the best in terms of RMSE as we have seen in the previous subsection. As the scatter scaling is increased, we notice that the PISLR approach does better for some intermediate scatter levels. But in the high-noise regime, the SSLR approach performs the best for most of the out-of-population cases, and the performance of the PISLR approach significantly degrades. We conclude that the SSLR features are the most robust for high scatter levels $\kappa$. 

We also show the relation between the order of the background scatter polynomial and the reconstruction accuracy in Figure~\ref{fig:rmse_vs_order}. For this experiment, we fixed the scatter scaling $\kappa$ to be $3$ and the standard deviation of the scatter kernel $\sigma_{\text{scatter}}$ to be $40$, which are both out-of-population values, and varied the background scatter polynomial order from 2 to 5. The mean RMSE across all test cases was noted for each order value and is plotted as a function of the order of the background scatter polynomial. The RMSE trend shows that the SSLR approach performs the best among all three compared schemes in this case. The PISLR approach provides slightly worse density errors than the SSLR approach.

\subsection{Effect of Gamma and Photon Noise Level on Reconstruction Accuracy}\label{sec:gamma_photon}
\begin{figure}
    \centering
    \begin{subfigure}[t]{0.45\textwidth}
        \centering
        \includegraphics[width=\linewidth]{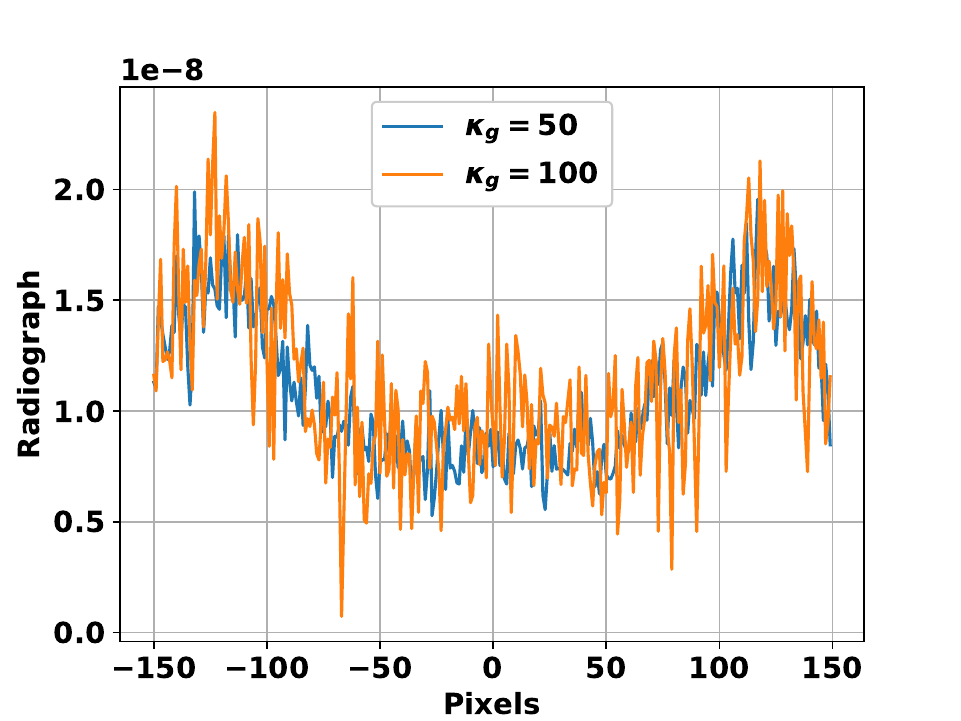}
        \caption{Scaling of Gamma Noise}
    \end{subfigure}
    \begin{subfigure}[t]{0.45\textwidth}
        \centering
        \includegraphics[width=\linewidth]{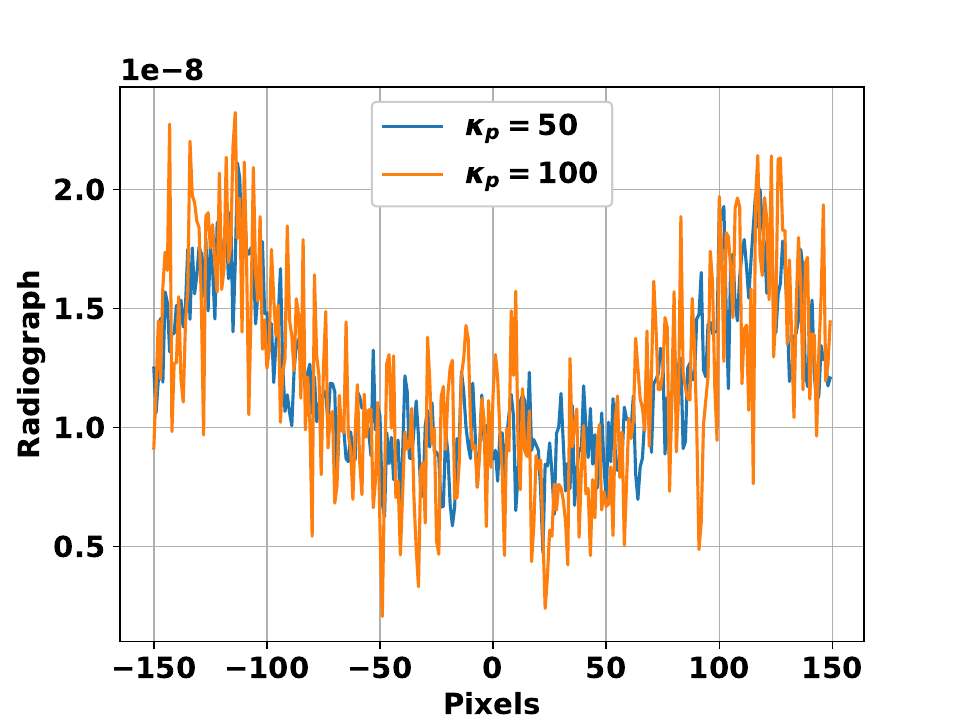}
        \caption{Scaling of Photon Noise}
    \end{subfigure}
    \caption{1D lineouts of noisy radiographs as a function of scaling of a) gamma and b) photon noise (as per Eq.~\eqref{eq:poisson_noise}). The other noise parameters used are $\sigma_{\text{scatter}}=40$, $\kappa=3$, and $n=2$.}
    \label{fig:noisy_rad_vs_gamma_photon}
\end{figure}

\begin{figure}
    \centering
    \begin{subfigure}[t]{0.5\textwidth}
        \centering
        \includegraphics[width=\linewidth]{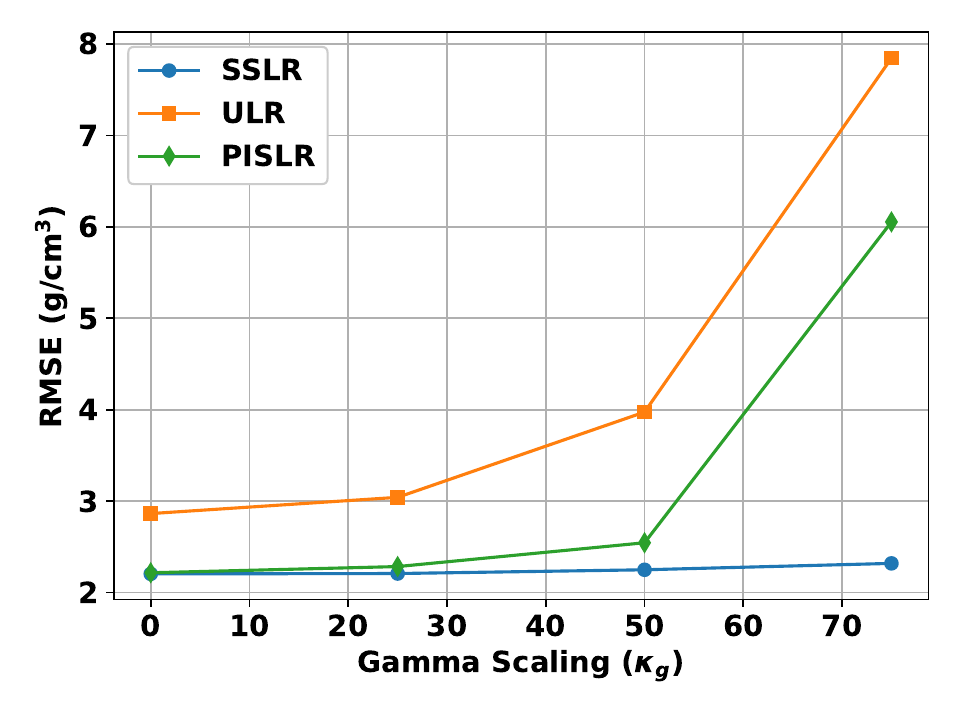}
        \caption{Scaling of Gamma Noise}
    \end{subfigure}%
    \begin{subfigure}[t]{0.5\textwidth}
        \centering
        \includegraphics[width=\linewidth]{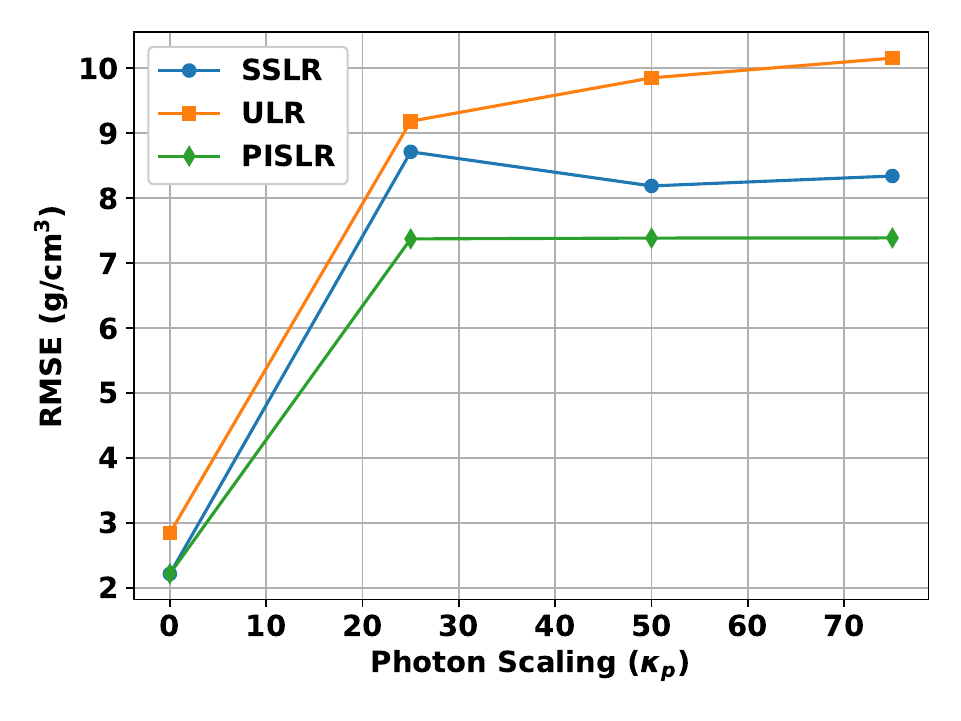}
        \caption{Scaling of Photon Noise}
    \end{subfigure}
    \caption{Variation of mean RMSE of reconstructed densities over all test cases as a function of scaling of a) gamma and b) photon noise. The other noise parameters used here were $\sigma_{\text{scatter}}=40$, $\kappa=3$, and $n=2$.}
    \label{fig:rmse_vs_gamma_photon}
\end{figure}

To evaluate the effect of the gamma and photon noise on the quality of reconstructed densities, we varied the amplitude and scaling of their respective kernels, i.e., $\kappa_g$ and $\kappa_p$ in Eq.~\ref{eq:poisson_noise}. The impact of increasing gamma and photon noise levels on the noisy radiographs is illustrated in Figure~\ref{fig:noisy_rad_vs_gamma_photon}.
The results in terms of RMSE values of reconstructed densities are summarized in Figure~\ref{fig:rmse_vs_gamma_photon} for noise parameters $\sigma_{\text{scatter}}=40$, $\kappa=3$, and $n=2$ (others as before). We can see that the SSLR approach is the most robust to high values of gamma noise, whereas, for high levels of photon noise, the PISLR approach remains the most robust. Importantly, both feature-based approaches outperform the {ULR approach} which has no explicit features.

\subsection{Comparison with Iterative Reconstruction Technique}
\begin{figure}[ht]
    \centering
    \includegraphics[width=0.6\textwidth]{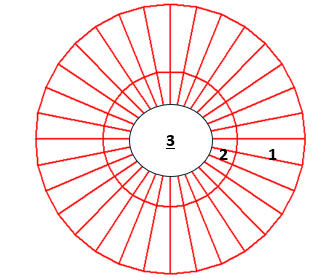}
    \caption{Regions utilized for model-based density reconstruction.}
    \label{fig:modelrecon}
\end{figure}

Finally, we compare our deep learning-based approaches with a traditional model-based iterative reconstruction (MBIR) method (Eq.~\eqref{eq:forward_model})~\cite{hanson1996bayes}. 
We consider an approach to reconstruct the density field where the domain is first divided into three regions as depicted in Figure~\ref{fig:modelrecon}.

In the first region, a model-based function form is prescribed based on characteristics of the underlying hydrodynamic behavior. Here we choose a model form with polynomial order $4$ in radial distance $r$ and linear in angle $\theta$ with 32 patches in the $\theta$ direction with continuity at each patch interface. A discontinuity is enabled at the shock location, i.e., a sharp interface between the first and second regions. The second region also employs a polynomial form for the density using the same order as the first region.  {Finally, within the third region, a pixelated model is utilized in close proximity to the gas-metal interface.} A traditional log smoothing prior (LogGGMRF) is utilized within this region~\cite{bouman1993generalized}. 
It is also noted that a model to differentiate the gas-metal is implemented based on the fact that gas has a maximum nominal density. 

The optimization starts with an initial guess for density {(taken to be close to nominal density)} in close proximity to the outer edge of region one. The scatter and gain are then obtained in the outer region, and the density in these regions is then subsequently optimized in conjunction with the scatter terms. Finally, optimization is performed using a non-gradient-based Brent method\cite{brent1973some}. This procedure optimizes the density and scatter up to the $2^{nd}$ region. After this optimization is performed, the mask is then extended into the shock region, whereupon the location and shape of the discontinuity are optimized.  Following this procedure, the density and shock location are simultaneously optimized, whereupon the optimization of the density in the second region proceeds up to the interior boundary. Finally, a pixelated density model is optimized to capture the higher frequency behavior of the Richtmyer-Meshkoff instability, with regularization provided by the LogGGMRF smoothing prior.  
This reconstruction procedure incorporates constraints for the conservation of object mass, a monotonicity constraint on the density field (from the outer edge to the shock), along with a smoothing constraint in the pixelated region interior to the gas-metal interface. 

\begin{figure}[ht]
    \centering
    \includegraphics[width=0.9\textwidth]{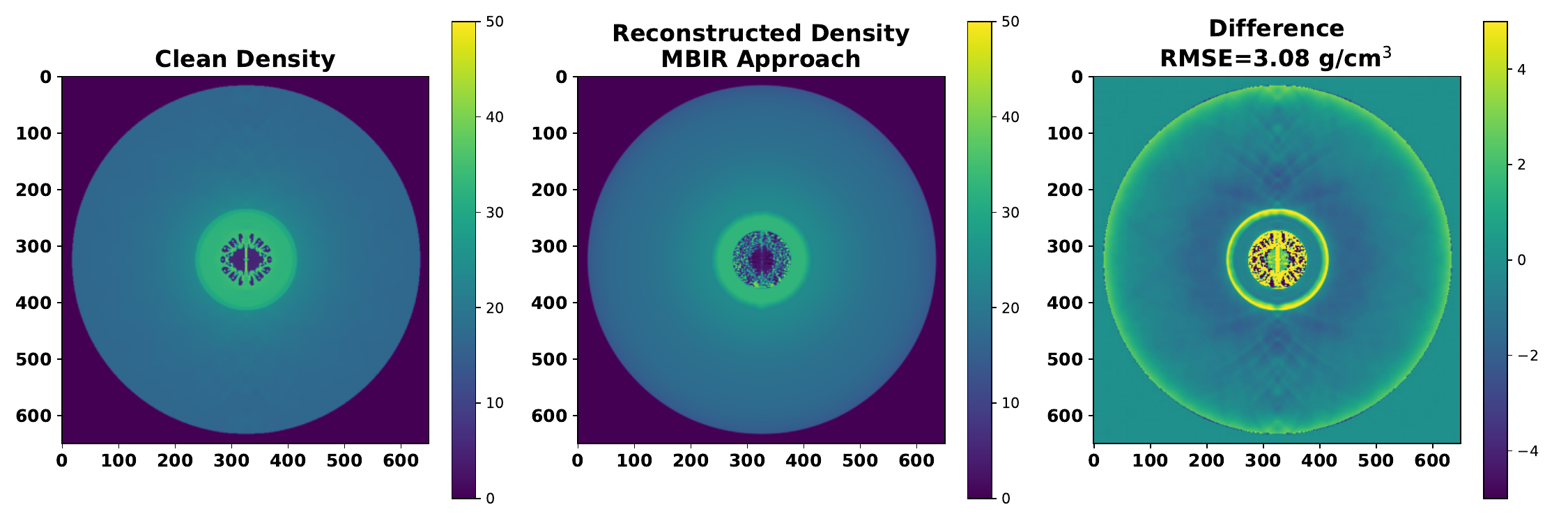}
    \caption{{2D profiles of the $32^{\textrm{nd}}$ frame for clean and reconstructed densities along with the difference (error) image using the MBIR method for in-population noise level. The RMSE of the reconstructed density is greater than that of the SSLR, ULR, and PISLR approaches discussed before.}}
    \label{fig:iterative_recon_2d}
\end{figure}

Since the optimization procedure for the MBIR approach with the scatter and noise model is computationally very expensive, we show the reconstruction results and comparison for an example density profile. In-population noise and scatter parameters were used to simulate the perturbations in this case. Figure~\ref{fig:iterative_recon_2d} presents the clean and reconstructed density using the MBIR method and the difference between the two images in $g/cm^3$. 

As may be observed from Figure~\ref{fig:iterative_recon_2d}, {the RMSE of reconstructed density ($3.08\,g/cm^3$) using the MBIR method is significantly larger than those obtained by the deep-learning-based approaches. The corresponding RMSE values of the SSLR, ULR and PISLR approaches for the same testing profile were $0.86\,g/cm^3$, $0.99\,g/cm^3$ and $0.88\,g/cm^3$ respectively}. In addition, accurate recovery of the 
gas-metal interface is not possible with the iterative scheme. Finally, if the functional form of the scatter model is uncertain (model mismatch), reconstruction is not possible without accurate knowledge of the underlying density field.

\subsection{Effect of Training Dataset Size}
In this subsection, we consider the effect of having limited training ground truth data (i.e., clean densities) on the accuracy of reconstructed densities using our proposed approaches. In other words, we explore the effect of changing the number of simulations used for training the models on the accuracy of reconstructed densities. For the sake of fair comparison, we kept the testing set fixed as in the previous case and trained all three approaches - ULR, PISLR, and SSLR on a much smaller training set (900 simulations). In Table~\ref{tab:rmse_vs_train_size}, we show a comparison of the RMSE values for the reconstructed densities using models trained with the smaller and bigger datasets. As can be seen from the table, the models trained on the bigger dataset perform better on the test set for both in and out-of-population perturbations. 
However, while the in-population performance of models trained on the smaller dataset is generally worse (by 0.12–0.22 g/cm$^3$ on average) for ULR and PISLR, the SSLR model shows a slight improvement in this case, though its out-of-population performance remains slightly worse. Nonetheless, for out-of-population data, SSLR approach's RMSEs show the least variation across training dataset sizes, indicating its robustness and generalization even with quite limited data.

\begin{table}[htbp]
\begin{tabular}{|c|cc|cc|}
\hline
Training Size & \multicolumn{2}{c|}{Bigger Set (9067 simulations)} & \multicolumn{2}{c|}{Smaller Set (900 simulations)}   \\ \hline
Approach Used & \multicolumn{1}{c|}{In-Population} & Out-of-Population & \multicolumn{1}{c|}{In-Population} & Out-of-Population \\ \hline
ULR    & \multicolumn{1}{c|}{\textbf{0.78 $\pm$ 0.24}}     &  2.84 $\pm$ 0.26  & \multicolumn{1}{c|}{\textbf{1.00 $\pm$ 0.24}}   &  3.07 $\pm$ 0.48       \\ \hline
PISLR     & \multicolumn{1}{c|}{0.90 $\pm$ 0.23}   &  2.21 $\pm$ 0.24 & \multicolumn{1}{c|}{1.02 $\pm$ 0.24}  &  2.35 $\pm$ 0.36    \\ \hline
SSLR   & \multicolumn{1}{c|}{1.21 $\pm$ 0.27} &  \textbf{2.19 $\pm$ 0.16 }   & \multicolumn{1}{c|}{1.03 $\pm$ 0.26}    & \textbf{2.23 $\pm$ 0.30}       \\ \hline
\end{tabular}
\caption{\changes{Comparing RMSE values of reconstructed densities for ULR, PISLR, and SSLR approaches for different training set sizes. Results are shown for in-population and out-of-population corruption levels. The values shown for each approach are mean $\pm$ std with units: g/cm$^3$. The numbers in bold indicate the lowest RMSE for each case. The parameters used for in and out-of-population noise are shown in table~\ref{tab:noise_params}.}}
\label{tab:rmse_vs_train_size}
\end{table}


\section{Conclusion} \label{sec:conclusion}
This work proposed a robust feature extraction-based approach to reconstruct hydrodynamic densities from measurements corrupted by noise and unknown scatter. We worked with 3D cone-beam CT radiographs of a sequence of ICF-like simulations. The ideal measurements were perturbed using simulated blur, scatter, and random noise. The encoder part of the proposed approach is learned to predict robust features from corrupted radiographs, and the decoder is then used to reconstruct densities from these features. Our results indicate that the networks trained to predict SSLR features as well as PISLR features prove to be most robust for density reconstruction in the presence of unknown or stochastic scatter and noise. In more extreme settings, the SSLR features offer the most robust approach.

\paragraph{Future Directions.}
Potential future directions could include comparing with adversarially trained robust networks.
This work can also be extended to feature-based robust reconstruction for other applications. \changes{Future work can also include the development of clean density or radiograph label-free unsupervised alternatives.}

\paragraph{Funding.}
This work was supported by the U.S. Department of Energy through the Los Alamos
National Laboratory (LANL) and the Laboratory Directed Research and Development program
of LANL.

\paragraph{Acknowledgment.}
The authors acknowledge Jennifer Schei Disterhaupt for her contributions to the noise and scatter model. The authors acknowledge Gabriel Malikal (PhD Student, MSU) for contributing to the code for radiograph generation for 3D cone beam CT. The authors also acknowledge Madeline Mitchell and Shijun Liang (MSU) for their work on the project.

\paragraph{Disclosures.}
The authors declare no conflicts of interest.

\paragraph{Data Availability Statement.}
The data underlying the results presented in this paper are not publicly available at this time but may be obtained from the authors upon reasonable request.

\bibliography{references}

\end{document}